\documentclass[twocolumn]{aastex63}

\usepackage{xspace}

\newcommand{\ang}[1]{#1\degr}
\newcommand{\eg}{\textit{e.g.}\xspace}
\newcommand{\ie}{\textit{i.e.}\xspace}

\received{January 30, 2020}
\revised{October 6, 2020}
\accepted{October 21, 2020}
\published{January 25, 2021}
\submitjournal{AstroPhysical Journal}

\shorttitle{Titan's haze seasonal variations}
\shortauthors{Seignovert et al.}

\begin{document}
\title{Haze seasonal variations of Titan's upper atmosphere\\during the Cassini Mission}

\correspondingauthor{Beno\^{i}t Seignovert}
\email{research@seignovert.fr}

\author[0000-0001-6533-275X]{Beno\^{i}t Seignovert}
\affiliation{Jet Propulsion Laboratory,
California Institute of Technology,
Pasadena, CA 91109,
USA}
\affiliation{GSMA, Universit\'{e} de Reims Champagne-Ardenne,
UMR 7331-GSMA,
51687 Reims,
France}

\author[0000-0003-0836-723X]{Pascal Rannou}
\affiliation{GSMA, Universit\'{e} de Reims Champagne-Ardenne,
UMR 7331-GSMA,
51687 Reims,
France}

\author[0000-0002-4320-2599]{Robert A. West}
\affiliation{Jet Propulsion Laboratory,
California Institute of Technology,
Pasadena, CA 91109,
USA}

\author[0000-0001-5541-2502]{Sandrine Vinatier}
\affiliation{LESIA, Observatoire de Paris,
Université PSL, CNRS, Sorbonne Université,
Université de Paris,
5 place Jules Janssen,
92195 Meudon,
France}

\begin{abstract}

This study presents a 13 years survey of haze UV extinction profiles, monitoring the temporal evolution of the detached haze layer (DHL) in Titan’s upper atmosphere (350-600 km). As reported by West et al. 2011 (GRL vol.38, L06204) at the equator, we show that the DHL is present at all latitudes below \ang{55}N during the northern winter (2004-2009). Then, it globally sunk and disappeared in 2012. No permanent DHL was observed between 2012 and 2015. It’s only in late-2015, that a new structure emerged from the Northern hemisphere and propagated to the equator. This new DHL is not as pronounced as in 2004 and is much more complex than the one observed earlier.
In one specific sequence, in 2005, we were able to investigate the short time scale variability of the DHL and no major changes was observed.
When both side of the limb were visible (dawn/dusk), we notice that the extinction of the DHL is slightly higher on the dawn side.
Additionally, during a polar flyby in 2009, we observed the longitudinal variability of the DHL and spotted some local inhomogeneities.
Finally, comparisons with UVIS stellar occultations and General Climate Models (GCMs) are both consistent with our findings.
However, we noticed that the timing of the DHL main pattern predicted by the GMCs can be off by up to \ang{30} in solar longitude.
All these observations bring new perspectives on the seasonal cycle of Titan's upper atmosphere, the evolution of the DHL and its interaction with the dynamics.

\end{abstract}

\doi{10.3847/1538-4357/abcd3b}

\section{Introduction}

Titan is the only moon of the solar system with a thick hazy atmosphere which represents approximately 20\% of its apparent
diameter. This atmosphere is mainly composed of nitrogen and methane. The photo-dissociation of these molecules by the
UV light in the upper part of the atmosphere leads to the production of a large number of other hydrocarbons and nitriles as
trace species and to photochemical haze. This haze is global and completely covers Titan. It controls
the thermal balance through its visible and thermal infrared properties \citep[\eg][]{Bezard2018}.
It also veils the lower atmosphere and the surface that can be perceived in a few methane windows in near infrared.

\begin{figure*}[ht!]
\plotone{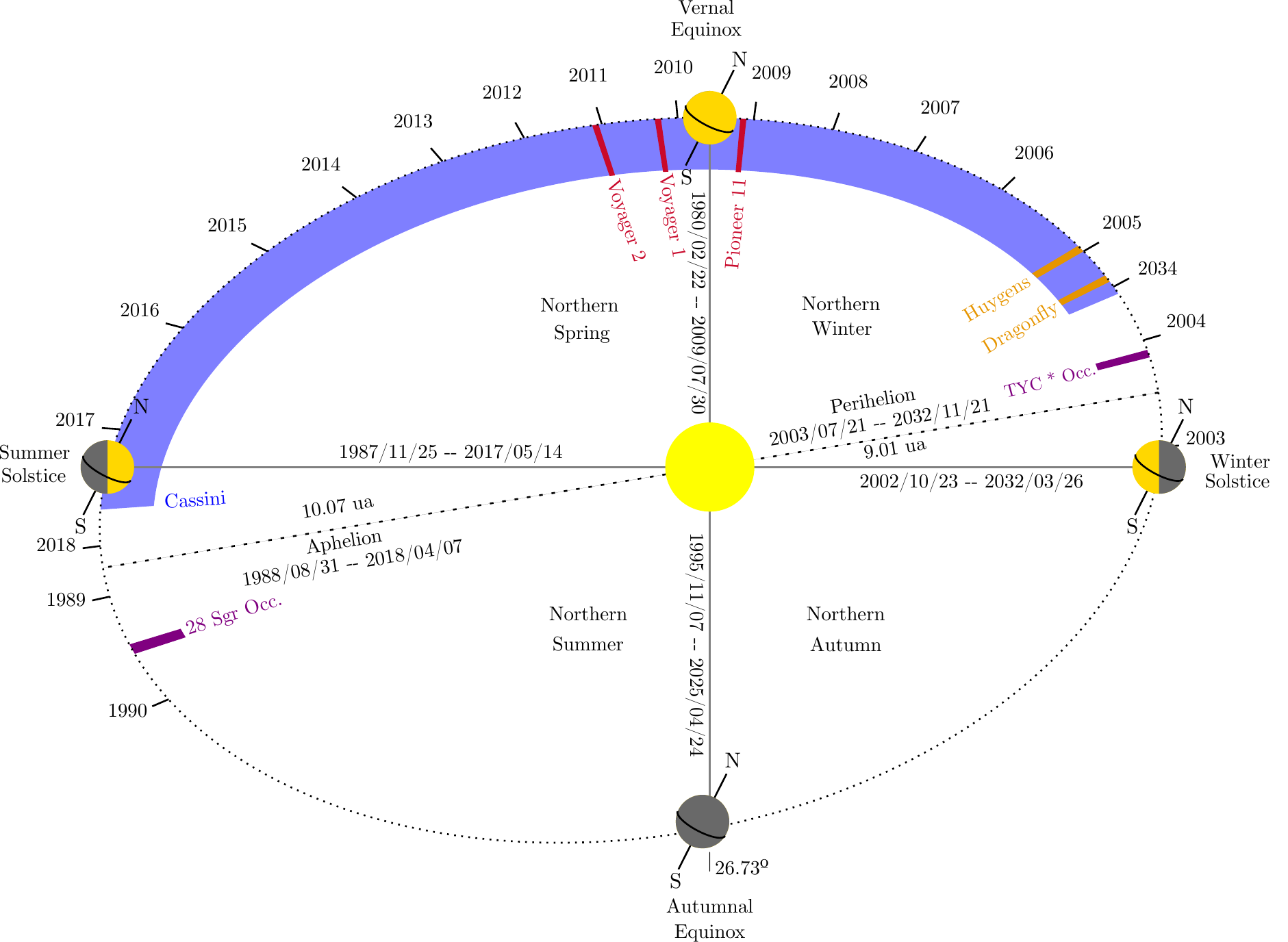}
\caption{Titan orbital position as a function of the season, reported as solar longitude position ($L_s$).
The Cassini mission covered almost half a Titan year. The Pioneer and Voyager flybys are also reported
as well as the Huygens landing and ground-based stellar occultations observed on Earth.}
\label{fig:titan_seasons}
\end{figure*}

Titan's haze was first resolved in the 80's by Pioneer 11 \citep{Smith1980} and the two Voyagers
\citep{Smith1981, Smith1982, Sromovsky1981}. It had several remarkable structures: a northern (winter)
polar hood, an interhemispheric asymmetry and a thin global detached haze layer (DHL) above the main global haze layer. It was
thought that the detached haze layer had a dynamical origin \citep{Smith1981}. Photometric analyses provided a means
to derive the extinction properties of both haze layers and to evaluate the effective radius of the aerosols in the detached
haze ($\simeq$ 0.3\micron) and in the main haze ($\simeq$ 0.4\micron) layers \citep{Rages1983, Rages1983a}. Analysis of Voyager images showed
that the detached haze layer appears due to a strong depletion of aerosol extinction around 300 km, yielding a distinct
layer above the main haze layer with a maximum extinction located around 350 km \citep{Rages1983}. Its horizontal extent was
very stable in pressure and it was reported at all the southern latitudes up to \ang{45}N where it connected to the northern
polar hood. The detached haze layer was re-observed twenty years after the Voyager flybys during Cassini first flyby in 2004
\citep{Porco2005}. The main change was in its altitude location at 500 km, which was 150 km higher than in 1981.
Again, it appeared as a fairly homogeneous global shell above the main haze layer at a constant altitude and
merged with the northern polarhood. Notably, while Voyager observations were performed after the northern spring equinox,
Cassini early observations occurred during the northern winter, that is half a season earlier (Fig.~\ref{fig:titan_seasons}).

 \cite{Toon1992} first attempted to explain the observation of the DHL. They used a 1D microphysical
model where an \emph{ad hoc} vertical wind maintains aloft the DHL particles at a constant altitude above the main
haze layer. Alternative scenarios were proposed to explain the DHL from purely microphysical processes. \cite{Chassefiere1995}
investigated the case of two different aerosol production layers. They proposed that the uppermost layer (500-1000 km)
produces fluffy aggregates that could be swept horizontally by winds, generating a detached haze layer. They also
proposed an alternative scenario where aerosols settle downward and interact with macromolecules from the main haze layer,
produced by the lower production zone (around 350-400 km). In the latter case, the interaction would produce by some
way an optical gap. However, they favored the scenario involving winds which would match all the constraints
known at that time. In the same vein, \cite{Lavvas2009} proposed a scenario based on a purely microphysical process. Aerosols
are produced at high altitude, as per the \cite{Chassefiere1995} hypothesis, growing as spheres down to levels around 500 km.
But there, the detached haze is produced by a sudden change in the fractal dimension of the aerosols. This produces a
sharp change in the microphysical properties, and an artificial optical gap. However, it is unclear how this model for the
production of the detached haze layer would be augmented to account for the seasonal evolution of the altitude,
disappearance, and reappearance (as described below).

\begin{figure}[ht!]
\plotone{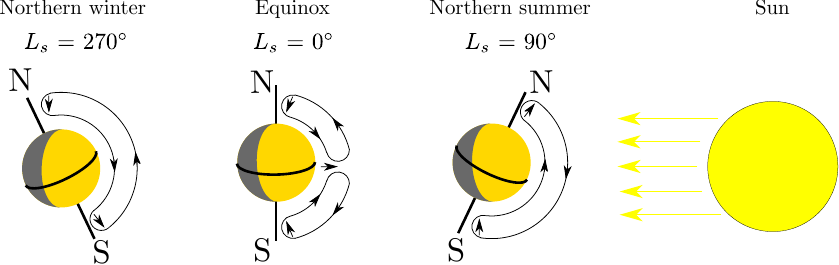}
\caption{Synthetic representation of Titan atmospheric circulation as a function of season.}
\label{fig:titan_atm_circulation}
\end{figure}

Later, with a 2D-General Climate Model (GCM) accounting for the transport of haze by dynamics and the radiative
feedback, it was possible to reproduce and explain the mechanism that produces the DHL \citep{Rannou2002}. It was also
demonstrated that this feedback strongly enhances the wind speed due to the thick polar haze cap near the winter pole.
In return, this cap enhances the cooling to space during the polar night \citep{Rannou2004} and
reinforces the circulation. Due to Titan's obliquity (\ang{27}) and the slow rotation rate, seasons are well
marked and Hadley circulation cells span both hemispheres. This situation leads to the formation of a
broad ascending circulation in the summer hemisphere able to lift aerosols up to high altitudes where they remain
suspended and are transported through mid-latitudes to the winter polar region where they are transported by
subsidence (Fig.~\ref{fig:titan_atm_circulation}). In this scenario, the location of the DHL corresponds to the area
where the settling speed is compensated by upward wind and evolves with seasonal changes of illumination.
More sophisticated 3D-GCMs improved the understanding of the haze cycle, including the formation of the detached haze,
and confirmed this picture \citep{Lebonnois2012,Larson2015}. This formation mechanism implies that the DHL
is a blending of aerosols newly produced and falling from above and older and larger aerosols produced in the
stratosphere and lifted by circulation.
Although the GCM results differ in some aspects with observations, they are able to capture the main picture behind
the existence of the DHL.

Photometric studies performed with Cassini data taken before the equinox in 2009 provide complementary observations
of the DHL \citep{Cours2011, Koskinen2011, Seignovert2017}.
On one hand, the authors used the light intensity scattered at the limb in UV (338 nm)
at different phase angles measured by ISS. On the other hand, a single value of the tangential opacity in VUV (187 nm)
was retrieved from UVIS observations during stellar occultations.
The results show the presence of large aerosols in the DHL with an effective bulk radius $\simeq$ 0.2\micron,
producing all the UV scattering, while small nanometric aerosols are needed to explain most of the VUV
extinction \citep{Cours2011}, which is quite consistent with a DHL made of two different populations of
aerosols.

\cite{West2011} also reported a rapid collapse of the detached haze layer starting just before the equinox. The altitude of
the DHL descended by about 80 km in 200 terrestrial days and by 30 km more in about 300 terrestrial days. A simple
extrapolation of the altitude of the DHL with time indicated that it would be at the same altitude as observed by Voyager
exactly one Titan year after the Voyager epoch. \cite{West2011} concluded that such a result was coherent with a seasonal
cycle of the DHL.  They compared their results with a 2D-GCM \citep{Rannou2002} and made a prediction about the reappearance of the
DHL several years later (2013-2016) at its initial altitude (around 500 km). \cite{Lebonnois2012} and \cite{Larson2015} made
similar predictions but with a reappearance of the DHL a bit later, around the next northern summer solstice ($L_S=\ang{95}$
and $\ang{70}-\ang{80}$, respectively). In practice, \cite{West2018} found that the DHL reappeared in early 2016
($L_S=\ang{73}-\ang{76}$) at 480 km, several months before the solstice (mid-2017). They followed the cycle of the DHL at the equator
and retrieved the haze extinction profile in the CL1-UV3 filter combination.
Its reappearance was much more complex than predicted. This early 2016 detached haze layer
dropped in altitude down to 470 km within a terrestrial year and vanished while a new DHL emerged again around 500 km. This
new layer appeared quite stable until the end of the Cassini mission (September 2017, $L_S=\ang{91}$). Unfortunately, no other
data exist to further probe the DHL and nothing is known about the fate of the detached haze after this date.

\medskip

In the present work, we perform a systematic latitude-altitude mapping of the detached haze layer in the range 350 to 600 km altitude.
This covers the period between July 2004 (half a season after the northern winter solstice) and the end of
the mission in September 2017 (after the summer solstice).
We used all the UV3 observations acquired by the Cassini Narrow Angle Camera (NAC) of ISS.
We used exactly the same model as \cite{West2018}, that is a ray tracing model in spherical shell geometry for the single-scattering albedo
and a correction for multiples scattering.

The outline of the article is as follows. In the next section (section~\ref{seq:observations}), we first give a global presentation of
the available data and the criteria we use to select images.
Then we describe the main principle of the retrieval model and the retrieval method.
In section~\ref{seq:lat_variations} we present the results of the photometric analysis as latitude -
altitude panels showing the spatial distribution of the DHL and the upper part of the main haze layer.
The seasonal cycle of the DHL is split in four specific periods between 2004 and 2017.
This section has then 4 subsections for each period where we explain in detail the main characteristics of the haze and
its evolution.
Section~\ref{seq:local_variations} is dedicated to the study of specific sets of observations that probe short time scales,
short term or diurnal variations. We first describe how the data were selected and then what they reveal about Titan's atmosphere.
In section~\ref{seq:comparisons}, we make comparisons between our results and results obtained at the same location and the same time with UVIS.
We also make comparisons between our results and prediction made by two Titan 3D-GCMs about the detached haze layer and its evolution.
The conclusion and the perspective of this work are given in section~\ref{seq:conclusions}.

\section{Observations and models}
\label{seq:observations}

\subsection{Selection of observations}
We conduct our survey on 138 images\footnote{See Data Availability section to get the list and the results for all the ISS images analyzed.}
taken by the Cassini Image Science Sub-System Narrow Angle Camera (ISS-NAC) with the
clear and ultra-violet filter combination CL1-UV3. At this wavelength (338 nm), ISS is sensitive to haze in the
upper part of Titan's atmosphere (300-600 km altitude). We choose the best sample among the 317 images available on the NASA Planetary Data System (PDS)
to get the highest temporal and phase coverage. For the main study we kept only images taken with at least a one-day gap.
We also kept specific sets of observations made a few hours apart to study short-term variations.

\begin{figure*}[ht!]
\plotone{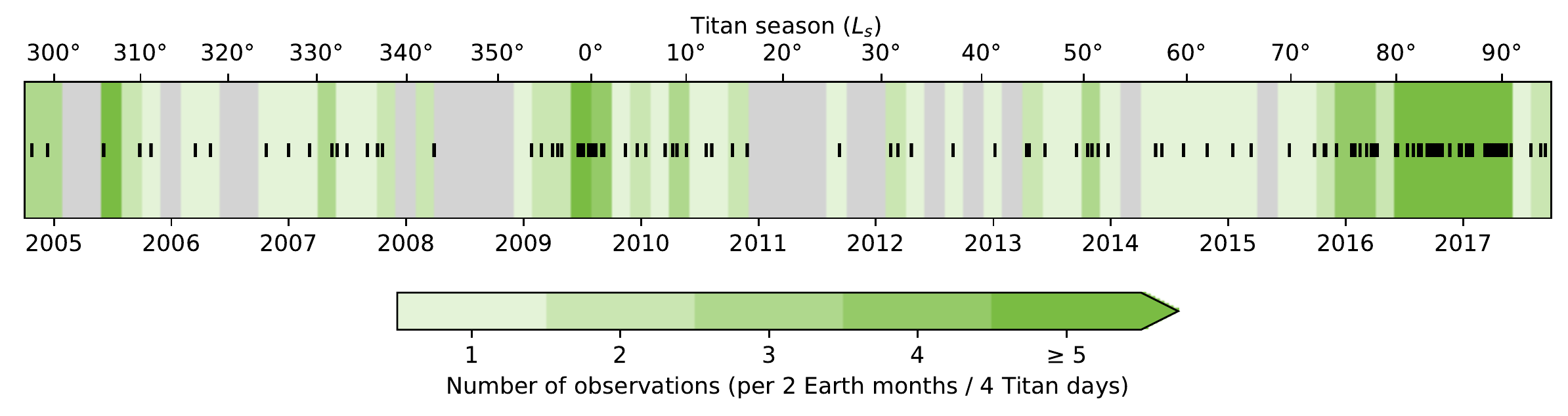}
\caption{Timeline of the ISS/NAC CL1-UV3 images analyzed (vertical black ticks). The number of observations per 2 Earth months period (\ie 4 Titan days / $\Delta Ls \approx \ang{2}$) is reported with a green gradient scale. The gray area correspond to the gaps with no available observations.}
\label{fig:img_sampling}
\end{figure*}

On average, the selected images are separated by 39 Earth days, \ie 2.5 Titan days (Fig.~\ref{fig:img_sampling}).
Although our sampling is not evenly distributed, due to orbital constraints and mission schedule, at least 90\% of the selected
images are separated by less than 120 Earth days, \ie 7.5 Titan days. Two main gaps of data can be observed.
The first is between 28 March 2008 and 25 January 2009 (302 Earth days/19 Titan days) and 26 November 2010 and 9 September 2011 (286 Earth days/18 Titan days).

\begin{figure*}[!ht]
\includegraphics[width=.4\textwidth]{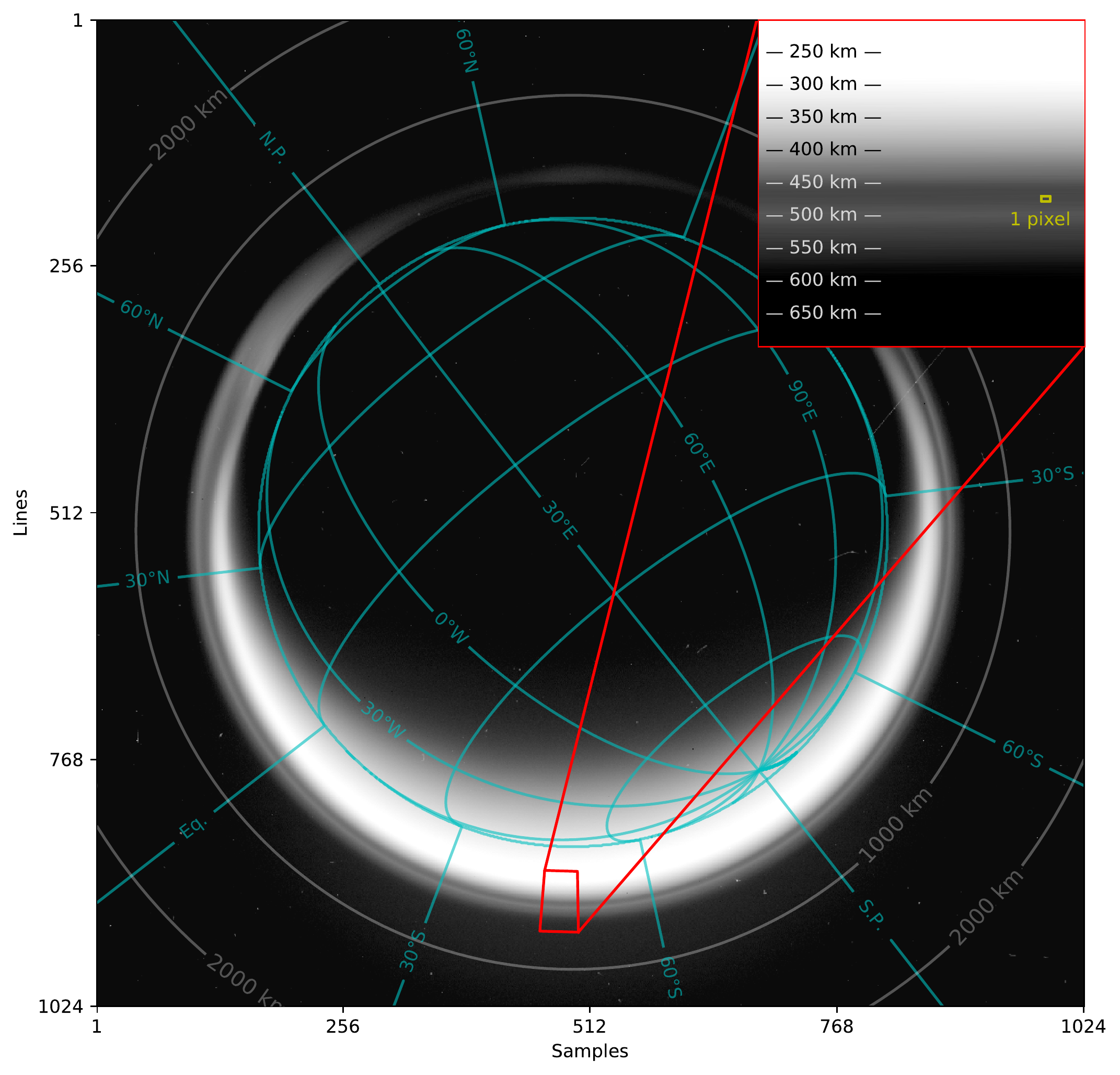}
\includegraphics[width=.57\textwidth]{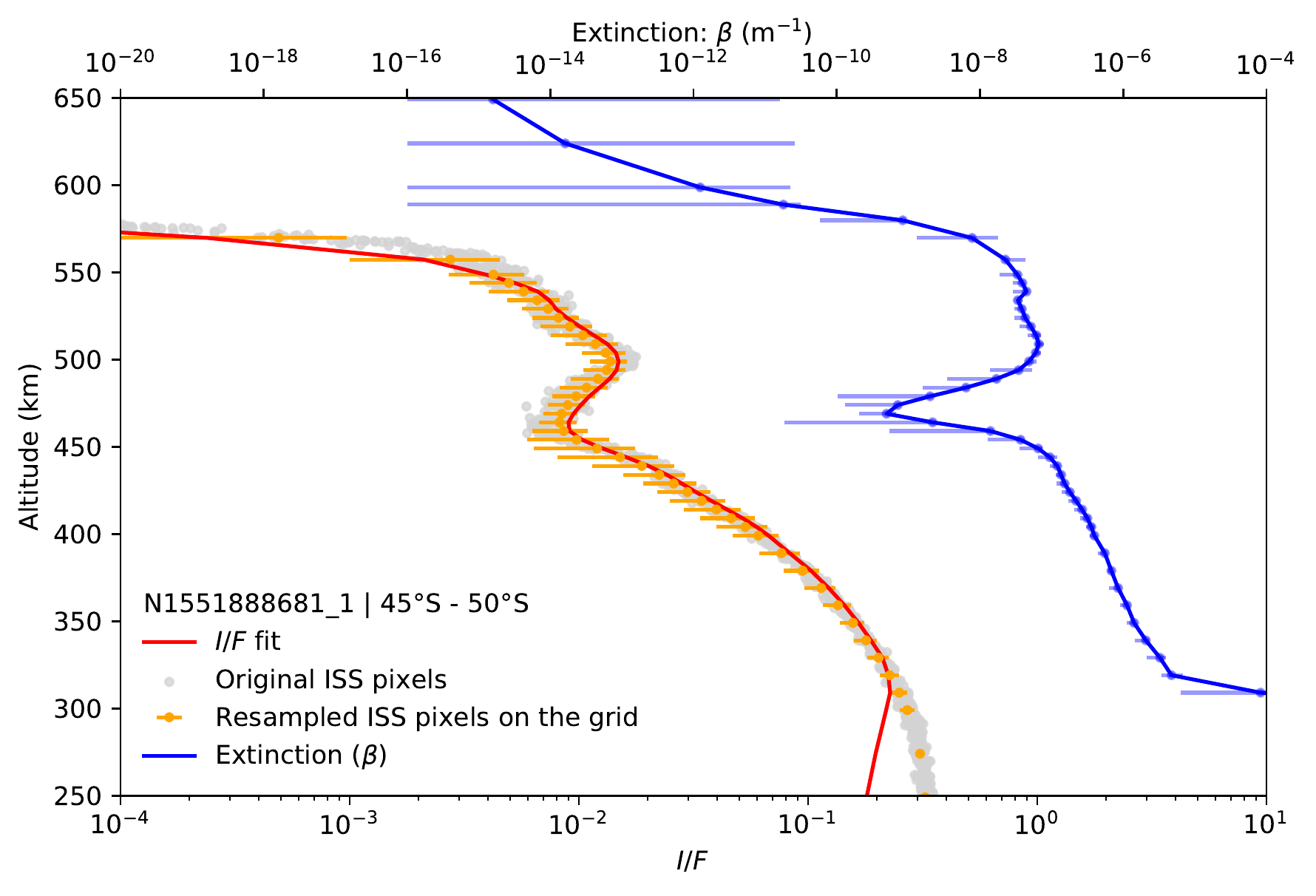}
\caption{Left: Log scale representation of the $I/F$  for the image
\textbf{N1551888681\_1} taken in March, 2007. The data are sampled at the limb by
bins of \ang{5}. The red area correspond to the bin at the photometric
equator ([\ang{45}S, \ang{50}S] and [\ang{76}W, \ang{79}W]). The pixel scale
(7.9 km) is represented in the zoomed area by the yellow rectangle.
An altitude grid is also represented where we can see the peak of intensity of the DHL
located at 500 km. Right: the red curve represents the inverted $I/F$ profile fitted
on the resampled data (orange error bars) within a \ang{5} bin in latitude.
The uncertainties on the retrieved extinction profile $\beta$ (in blue) is calculated to match the $1 \sigma$ distribution of the $I/F$ pixels. The model fits the observations down to 300 km.
The uncertainties increase rapidly when observed ($I/F$) is close
to the noise level ($3\cdot10^{-3}$).}
\label{fig:model_uncertainties}
\end{figure*}

The selected images are calibrated using the CISSCAL routines (v3.8) provided on the Planetary Data System. To improve
the signal to noise ratio on the limb profile, we deconvolved the images with a Poisson Maximum A Posteriori
(PMAP) method using the point spread function (PSF) measured in-flight \citep{West2010, Knowles2020}.
This deconvolution method is known to be efficient to restore fine strutures in astronomical images \citep{Hunt1996}.
In our case, the main source of light is always Titan and is resolved inside the field of view of the camera (Fig.~\ref{fig:model_uncertainties}a). Therefore, we don't expect to see a significant contribution of the stray light in this configuration \citep{West2010}.
The image pointing is initialized with the SPICE kernels \citep{Acton1996, Annex2020}. Since the surface of Titan is not visible in the UV, we improved
the location of Titan's center by fitting the limb intensity. Then, we calculated the planetocentric coordinates of each
pixel and their tangent point altitude with respect to a mean spherical body with a radius of 2575 km.

Intensity profiles are extracted every \ang{5} on both sides of the limb. Depending on the
latitude of the Sub-Cassini point on the ground, the sampling in latitude is not evenly distributed for each image.
Therefore, the polar latitudes are usually less covered than the equator.
Also, the solar illumination changes drastically during the
season between the northern mid-winter to summer, which restricts our ability to see both poles at the same time.
On average, the image pixel scale is about $10 \pm 5$~km.

\subsection{Model of scattering at the limb}

To retrieve the haze extinction profiles from the intensity observations, we model the synthetic
radiance factor ($I/F$) with a single scattering ray tracing model in a spherical shell geometry.
The effect of multiple scattering is accounted for as a correction $\varrho_k (z)$
applied to the volume scattering along line of sight.
This technique was used successfully several times before \citep[\eg][]{Rages1983, Rannou1997, Seignovert2017, West2018}.

In the detached haze layer, the multiple scattering is mainly produced by the light coming from the main haze below. To
evaluate $\varrho_k$, we use a representative vertical profile of the main haze that
reproduce the observed intensity of Titan in the UV. With a radiative transfer model \citep[SHDOMPP, from ][]{Evans1998}
we have access to the complete radiative source function at each level of the atmosphere as well as the optical properties (haze absorption and scattering and Rayleigh scattering). We are then able to compare the
intensity that is scattered in the direction of the observer from the direct sun only and from the direct sun and the scattered
field coming from below. $\varrho_k$ is defined as the ratio of multiple scattering to single scattering toward the observer
for a given altitude and as a function of the incident and emergent angles. This parameter is pre-computed as a function of
altitude, incident and emergent angles and saved in a look-up table \citep[see.][for details]{West2018}.
We find that the multiple scattering increases the scattered intensity at the limb of Titan in the UV by a ratio between
1.05 and 1.20, depending on the geometry of the observation.
This effect is included in our model. We also checked that significant changes in the main haze (single scattering albedo, the opacity and the vertical scale height) only affect the value of $\varrho_k$ by a few percent and can be neglected.

We discretize the atmosphere in $N = 60$ irregular layers of various thickness : $\Delta z =$ 50 km from the
ground to 200 km, $\Delta z =$ 25 km from 200 to 300 km, $\Delta z =$ 10 km from 300 to 400 km and from 550 to 700 km.
Finally, we used $\Delta z$= 5 km between 400 and 550 km. This grid allows us to take advantage of the spatial resolution
of the ISS NAC camera in the region of interest where is mainly located the DHL.

We can write the outgoing $I/F (z)$ as:

\begin{equation}
I/F (z) = \sum_{i=0}^{n_x-1} \int\limits_{x_k}^{x_{k+1}}
\frac{\left< \varpi P(\Theta) \right>_k}{4}
e^{-\left( \tau^i_k\left(z\right) + \tau^e_k\left(z\right) \right)}
\beta_k\left(z\right) \varrho_k\left(z\right) d{x}
\label{eq:west2017_sup_limb}
\end{equation}

where the summation is performed on the $n_x-1$ segments defined by the intersections of the line of sight and the spherical shell boundaries. The impact parameter $z$ (the lowest altitude reached by the line of sight) is given by the bottom of the $n^\mathrm{th}$ layer crossed. Therefore, each layer of the atmosphere is crossed twice. $x$ is the abscissa along the line of sight. $\tau^i_k$ and $\tau^e_k$ are the opacities along the incident and emergent paths.
$\left< \varpi P(\Theta)\right>_k$ is the average of the product of the single scattering and the phase function at the scattering angle of the observation $\Theta$ for the layer crossed on $x_k$.  $\beta_k(z)$ is the local extinction coefficient at altitude $z$ (and the product of the haze cross-section and the local number density). Here, the altitude $z(x)$ is the local altitude at point of abscissa $x$ along the line of sight.

\subsection{Retrieval method}

Based on our previous work \citep{Seignovert2017, West2018}, we make the assumption that the optical properties $\left<\varpi P(\Theta)\right>_k$ of the aerosols are constant in the upper part of Titan's atmosphere. This allows us to focus our study only on the retrieval of the extinction along the line of sight.
In our model, the $I/F (z)$ intensity profiles depend on the set haze extinction profile $\beta(z)$ and the viewing geometry of the observation (incidence, emergence and phase angles).
We assume no horizontal in-homogeneity along the line of sight \citep{Seignovert2017}. We retrieve a set of extinction values $\beta_i$ (the vector ${\beta}$), with $i$ the indices of the layers, that matches the values of the $I/F_i$ (the vector $I/F$).

From the Eq. (\ref{eq:west2017_sup_limb}), it is possible to cast the scattered intensity $I/F_i$, as a function of the extinction $\beta_j$ with $j \le i$. This forms a non-linear triangular system. To find the vector $\beta$, we have to solve the formal equation:

\begin{equation}
    I/F = G(\beta)
\end{equation}

where ${G}$ is a nonlinear function which depends on $\beta_i$ and on the viewing geometry of the observation.

We solve the system by minimizing globally the difference between the modeled $I/F$ and the observations
using a Levenberg-Marquardt minimization. Therefore, we obtain simultaneously all the $\beta_i$ at once.
Moreover, the tangential opacity along a line of sight ($\tau_{los}$) is considered opaque when it reaches 3 (usually around 300 km in the UV).
Beyond this threshold, we do not retrieve the value of $\beta$.
An example of inversion is presented in the figure~\ref{fig:model_uncertainties}.

The uncertainties associated to the intensity profile ($I/F$) correspond to a $1 \sigma$ distribution of the observed $I/F$ for a given altitude within a \ang{5} bin in latitude.
The number of pixels within a \ang{5} bin varies from one image to another and depends on the viewing geometry and the altitude sampled. Usually we have more than 1500 pixels per profile between 0 and 700 km.
In the worse case scenarios (N1630432142\_1 at $6 \sigma$ of the average pixel scale) we have at least 120 pixels per profile.
This allow us to resampled our observed profile on a vertical grid with layers smaller than the image pixel scale in the 500 km region.
The uncertainties in the extinction profile ($\beta$) correspond to the minimum and maximum values required to fit the data within $1 \sigma$ of the observed $I/F$ in each layer independently \citep[contrary to][ where the errors were computed in all the layers simultaneously]{West2018}.

\section{Seasonal cycle of the haze extinction}
\label{seq:lat_variations}

In order to provide a detailed explanation of the complex latitudinal variability of the detached haze layer, we present
some of the key images that we have analyzed. Based on our previous work, focused on the evolution of the detached haze in the
equatorial region \citep{West2018}, we define four different seasonal phase characteristics of the evolution of the detached haze layer.
Between 2004 and 2008, the DHL was stable in altitude and extinction profile.
Between 2008 and 2012, it settled and disappeared in the main haze below the 300 km.
During the period 2012-2016, the DHL was not observed and only sporadic transitory layers showed up in the UV3 images.
After 2016 and up to the end of the Cassini mission, the DHL reappeared following a complex pattern.
Here we illustrate the complete time and latitude survey, covering a period of time which includes about half a Titan year.
This is valuable because it encompasses the equinoctial transition period of 2009.
For each phase, we display the altitude and latitude distribution of the instantaneous haze extinction coefficient retrieved from intensities at the illuminated limb of Titan.
The same color scale is applied to all the panels in order to keep a consistent view on the whole data-set.
It corresponds to the range where the model have the best sensitivity at the UV wavelengths. In some cases, smaller features might be harder to highlight with this color scheme, however, all the results from theses panels are available to the reader for more in depth considerations (see Data Availablity section).
Locations where no data are available are left as blank areas on the panels.

\subsection{Period 1: A very well delimited detached haze Layer during the Northern Winter (2004-2008) - $L_s=\ang{300}-\ang{340}$}

\begin{figure*}[!ht]
\plotone{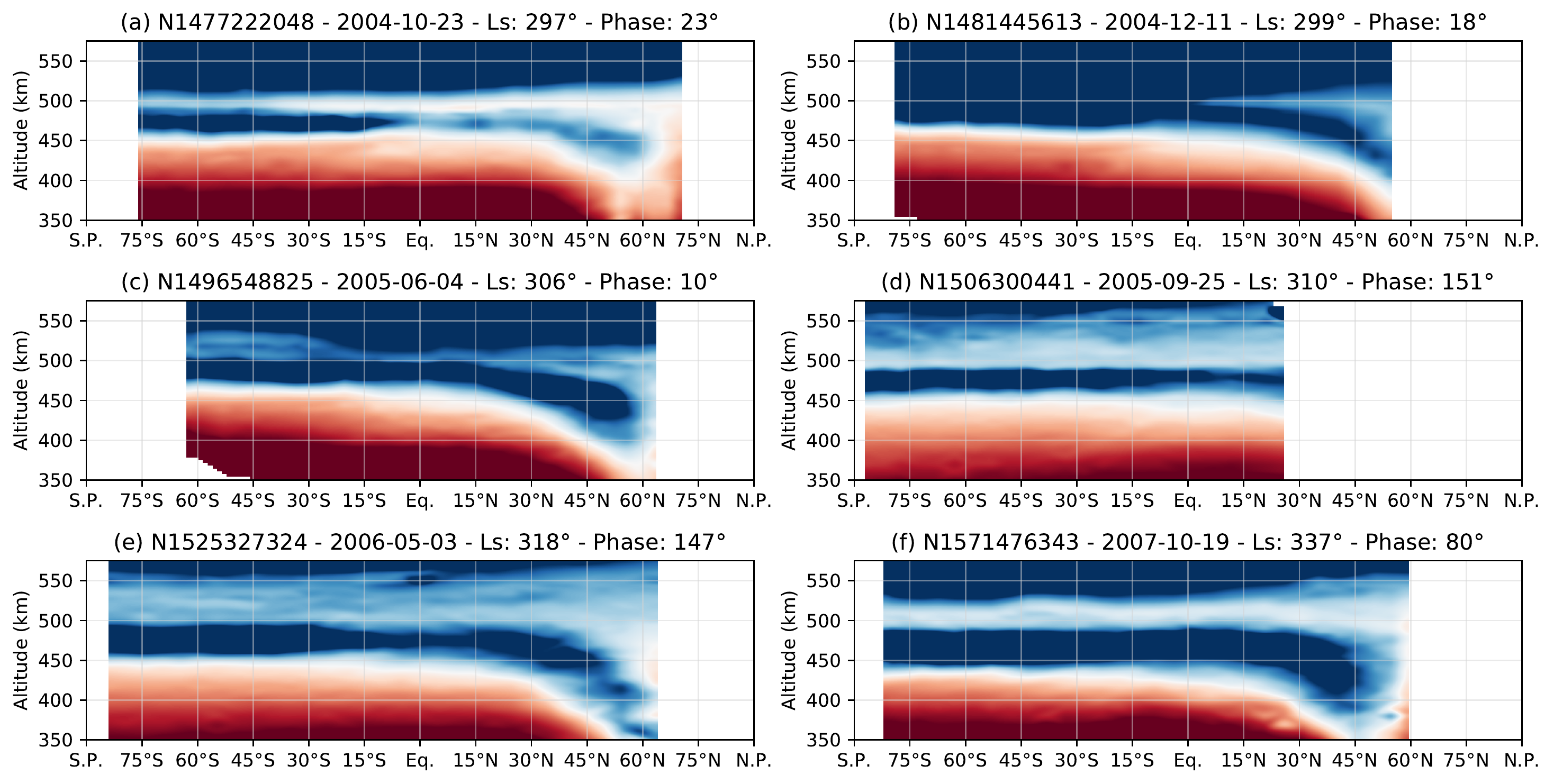}
\includegraphics[width=.5\textwidth]{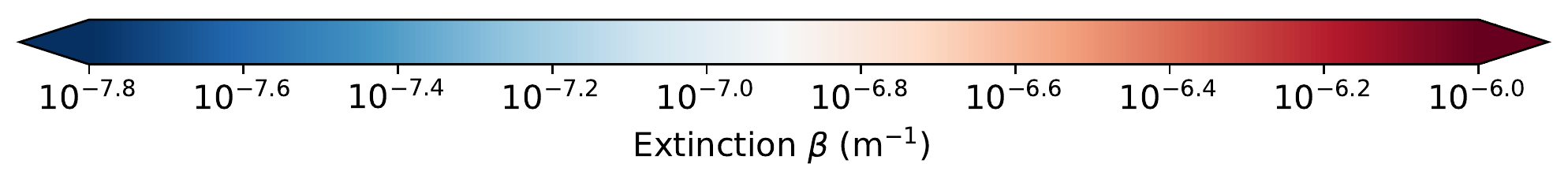}
\caption{Latitudinal haze extinction profile ($\beta$) retrieved for 6 images taken between 2004 and 2008
($L_s=\ang{300}-\ang{340}$) showing a stable DHL at $500 \pm 20$ km altitude.
The color schema is fixed for all the figures to make direct comparison
between the different panels. The seasonal solar longitude ($L_s$) and the observation phase angle are
also provided in each panel.}
\label{fig:dhl_2004_2008}
\end{figure*}

At its arrival in the Saturnian system in 2004, Cassini observed a single detached haze layer at 500 km altitude
(Fig.~\ref{fig:dhl_2004_2008}a) similar to the one observed at 350 km by Voyager 24 years before
\citep{Smith1981}. At that moment, Titan was two years after the winter solstice in the northern hemisphere, at $L_s=\ang{300}$.
With Cassini, we see that, in the southern hemisphere, the haze layer was completely detached from the main haze layer.
The haze extinction was at least one order of magnitude smaller inside the depleted zone (470 km) than in the main and
the detached haze layers (below 450 and at 500 km respectively).
Between the equator and up to about \ang{60}N it presented a local depletion in extinction
of a factor 10. There, the separation with the main haze is not as distinct as in the south, but is still sufficiently significant to
defined a detached haze layer.
The altitude of the depletion zone decreased by about 50 km between latitude \ang{30}N  and \ang{60}N.
The detached haze layer merged with the polar hood beyond \ang{60}N. This description of the detached haze layer at
the beginning of Cassini mission is very consistent with the results obtained from stellar occultation in 2003 \citep{Sicardy2006}.
Throughout the period 2004-2008 the detached haze layer was quite stable in shape and  altitude, with a maximum of extinction
at $500 \pm 20$ km. The top of the main haze layer was located around $450 \pm 20$ km below \ang{30}N and dropped
by 50 km between \ang{30} and \ang{60}N.

However, there are noticeable variations in haze extinction. During several months, the
detached haze remained stable, but in December 2004 (Fig.~\ref{fig:dhl_2004_2008}b),
the detached haze layer extinction was found to be a factor of 10 lower than previously at almost all latitudes south
of \ang{30}N, and about half a decade above \ang{30}N. The polar hood and the main haze do not show a similar decrease.
In the following periods, (Fig.~\ref{fig:dhl_2004_2008}c, d and e),
the detached haze was partially restored, but not with the same amount of extinction as before. Only observations in
2007 (Fig.~\ref{fig:dhl_2004_2008}f) show extinctions in the detached haze layer comparable to those seen before
the decrease. We note that the decrease of extinction below 370 km in the polar hood above \ang{50}N
(Fig.~\ref{fig:dhl_2004_2008}e) is at the limit of sensitivity of the UV filter. The stability of the large-scale
structure of the detached haze layer is related to the steady state of the large-scale circulation during all the winter.
The observation of October 2007 (Fig.~\ref{fig:dhl_2004_2008}f) is the last view that we have of this stable state
before the seasonal turnover.

During this period, the detached haze also has a strong layering with, at some latitudes, distinct decks which are not
continuous and rather appear as foliation. This feature is more pronounced in some observations, for instance from
June 2005 to May 2006, but does not shows up in October 2007, except marginally beyond \ang{30}N. The foliated detached
haze layer has a larger geometrical thickness than before December 2004.

As we will see in the section~\ref{seq:local_variations}, the detached haze layer exhibits some longitudinal or diurnal variability
that limits our ability to interpret details of features observed in single images. The small-scale features
could depend on the local short-term dynamics such as initia-gravity waves.

\subsection{Period 2: Drop and disappearance of the main haze layer around the Vernal Equinox (2008-2012) - $L_s=\ang{340}-\ang{30}$}

\begin{figure*}[!ht]
\plotone{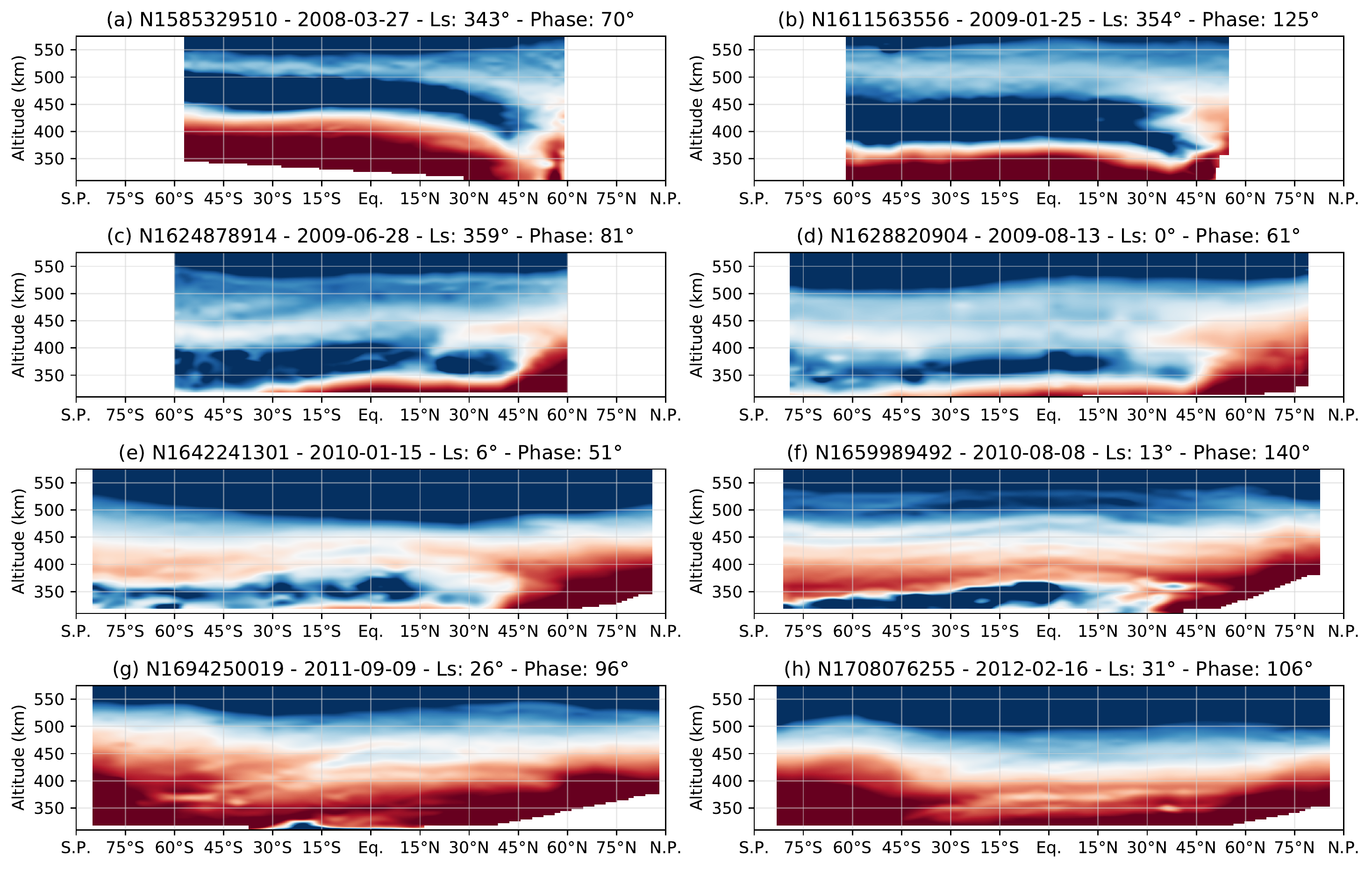}
\includegraphics[width=.5\textwidth]{Extinction_colorbar.pdf}
\caption{Same as the figure~\ref{fig:dhl_2004_2008} for 8 images taken between 2008 and 2012
($L_s=\ang{340}-\ang{30}$) showing the drop and disappearance of the DHL.
The color schema is identical to the figure~\ref{fig:dhl_2004_2008} to provide
direct comparisons. In this case, the altitude range is extended down to 300 km
(where the model is less reliable).}
\label{fig:dhl_2008_2012}
\end{figure*}

A precursor sign of the drop of the detached haze can be seen in March 2008 (Fig.~\ref{fig:dhl_2008_2012}a).
The main haze starts an initial contraction around \ang{35}S. There, the depleted zone is almost
75 km thick at its maximum. In January 2009, the main haze continued to fall from 425 km down to 375 km
while the detached haze layer remained around 500 km (Fig.~\ref{fig:dhl_2008_2012}b). After the drop
of the main haze in early 2009, the detached haze starts its own descent in June 2009, just before the equinox
(Fig.~\ref{fig:dhl_2008_2012}c). This delay in collapse increased the apparent thickness of the depletion
zone between the two haze layers.

As for the main haze, the detached haze collapses first in the Southern hemisphere, from 500 km to 425 km, and
then at the equator and in the northern hemisphere (Fig.~\ref{fig:dhl_2008_2012}c).
This is associated with the circulation turnover affecting first the summer hemisphere ascending branch.
With time, the detached haze gradually settled in altitude and finally disappear below 300 km.
Later ISS observations made with the Blue and Green filters visually show that the main DHL continues
its descend below 300 km during 2011. Since our current model was only tested for UV observations, we were not able
to observe its merge with main haze.
The complete collapse of the detached haze, as it appeared in the UV3 filter, is displayed in Fig.~\ref{fig:dhl_2008_2012}c to
Fig.~\ref{fig:dhl_2008_2012}h. We note that the column extinction is smaller at equator than at
other latitudes, and this is the case during entire period of the collapse.

During the fall, a second thin detached haze layer, at planetary scale, is evident above the collapsing detached
haze layer. In January 2010 (Fig.~\ref{fig:dhl_2008_2012}e), the detached haze layer was located between 375 and 400 km.
We can still see a double deck of haze, and this time the detached haze appears higher at the equator compared to the two
hemispheres, producing an arch. The haze peak extinction has globally increased by a factor of two due to sedimentation
in denser layers.

In August 2010 (Fig.~\ref{fig:dhl_2008_2012}f), one year after equinox, the detached haze layer continued
its drop down to 375 km around \ang{40}S and 400 km at the equator. It has gained in complexity with
multiple secondary layers up to 520 km. The detached haze formed a remarkable arch with a difference of about 50 km
in altitude between the equator and the poles as previously noticed by~\cite{West2011}.
This observation and the next one correspond to the same seasonal phase  of Voyager flybys ($L_s=\ang{8}$ and \ang{18}).
They can be compared directly.
We now know that this season was a time of rapid change, and that the Voyager probes observed transient situations.
Voyager also observed the detached haze higher near equator than elsewhere \citep{Rages1983, Rannou2000}.

Due to orbital constraints and mission planning, the next observation was made in September 2011
(Fig.~\ref{fig:dhl_2008_2012}g). The detached haze layer was, at that time, well below the level of the polar hoods.
Again, secondary detached layers show up as high as 470 and 520 km.

The south polar hood was not present in August 2010 (Fig.~\ref{fig:dhl_2008_2012}f) and appeared in less than 13 months.
This indicates that the circulation started to reverse around the equinox and the southward circulation sent haze to
the southern polar region and produced a polar hood. The change in haze distribution
is a very good indication of the timing of the equinoctial circulation turnover, as  discussed later. We note
that the strong haze depletion at 300 km and between \ang{30}S and \ang{20}N is real (and visible in the I/F profiles) but may be exaggerated at \ang{20}N
due to the limit of the retrieval procedure. At this altitude level, Titan's atmosphere is opaque to UV radiation
(see Fig.~\ref{fig:model_uncertainties}) and does not allow us to follow the main depletion below this altitude.

The last UV3 image we have showing a detached haze layer was taken in February of 2012 (Fig.~\ref{fig:dhl_2008_2012}h). At that
time, the initial detached haze had completely disappeared and the secondary detached haze layer was still descending
and had reached 400 km altitude. The secondary detached haze layer is not well delineated by a layer strongly depleted in aerosols.
The south polar hood increased its latitudinal extent northward to \ang{50}S and became larger than the northern
polar hood which tends to retreat.

\subsection{Period 3: Absence of the DHL, with sporadic transitory layers after the Northern Spring Equinox (2012-2015) - $L_s=\ang{30}-\ang{75}$}

\begin{figure*}[!ht]
\plotone{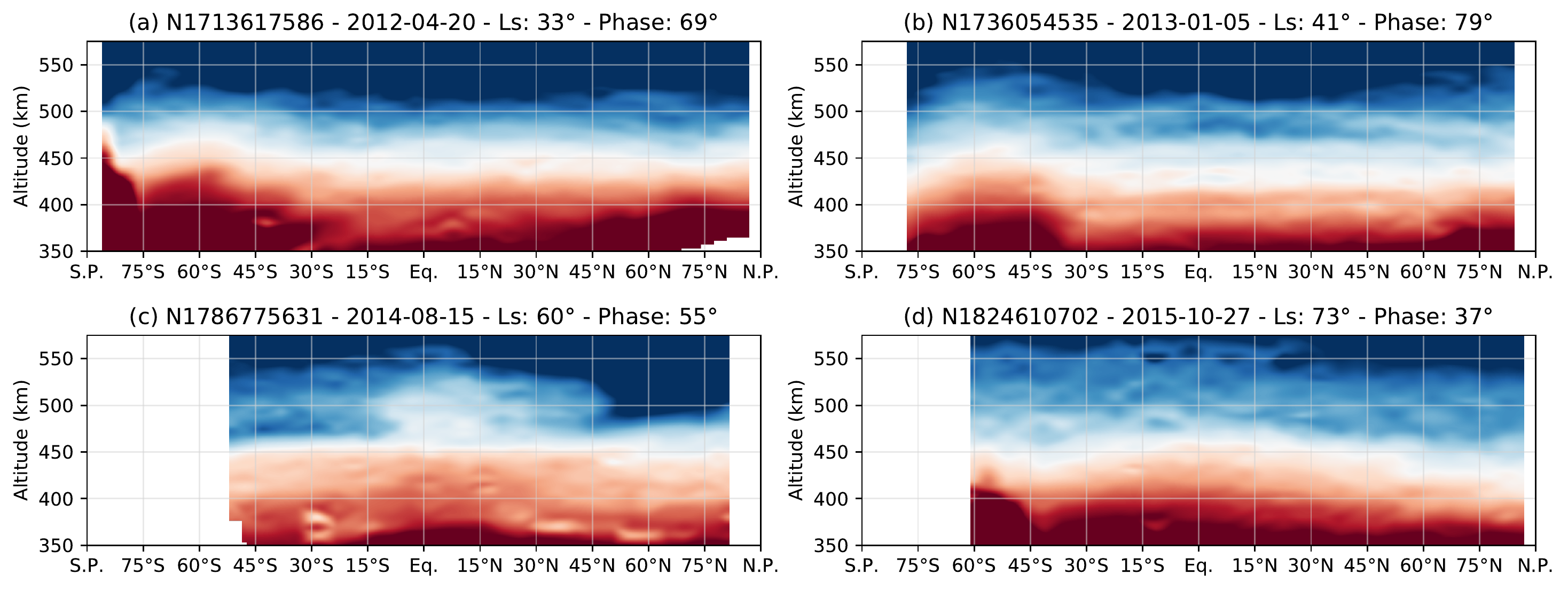}
\includegraphics[width=.5\textwidth]{Extinction_colorbar.pdf}
\caption{Same as the figures~\ref{fig:dhl_2004_2008} and~\ref{fig:dhl_2008_2012}
for 4 images taken between 2012 and 2015 ($L_s=\ang{30}-\ang{75}$) showing sporadic
transitory layers and the absence of a stable DHL.
The color scheme is the same as in previous figures and the altitude extends down to 350 km.}
\label{fig:dhl_2012_2015}
\end{figure*}

During this period (Fig.~\ref{fig:dhl_2012_2015}), the main haze layer has large-scale structures which slowly evolve under the influence of the
large-scale circulation. The south and north polar hoods are still visible and they evolve
with time. Superimposed on this background haze, transient structures show up and disappear from one observation
to the other. At some moments, large-scale detached hazes appear. They differ from the detached haze seen at the
beginning of the mission because they are not stable in time and in altitude.

In April 2012 (Fig.~\ref{fig:dhl_2012_2015}a), the detached haze has completely disappeared, except some residual
structures around the equator at 370 km around the equator.
These relics of the last detached haze layer are almost not perceptible in the corresponding I/F profile. At other
latitudes, we can see only the main haze with a marked south polar hood and a small increase of extinction above
\ang{65}N that could be the residual north polar hood. Sometimes, detached layers emerge from the background with large latitudinal
extent (e.g. detached haze at 500 km Fig.~\ref{fig:dhl_2012_2015}b). However, they only remain for a short time
and are not seen in the following observations.

In August 2014 (Fig.~\ref{fig:dhl_2012_2015}c) we observed a plume of aerosol between \ang{10}S and \ang{25}N,
reaching 530 km. A detached haze layer seems to spread from this plume toward the north and the south. This
detached haze is around 500 km altitude, descending to 470 km at \ang{50}S (and probably even further south). In the
north, the detached haze does not extend further than \ang{50}N and remains at 500 km. This indicates an
atmosphere circulation flowing from equator toward the south pole. These aerosols seem to originate
from the equatorial part of the main haze.

Most of the observations between 2012 and the end of 2015 are featureless as in figure~\ref{fig:dhl_2012_2015}d)
taken in October, 2015. During this period, the main haze has a uniform scale height of 45 km and
with an homogeneous extinction at the planetary scale.

\subsection{Period 4: Reappearance of a new and weaker detached haze layer around the Summer Solstice (2015-2017) - $L_s=\ang{75}-\ang{95}$}

\begin{figure*}[!ht]
\plotone{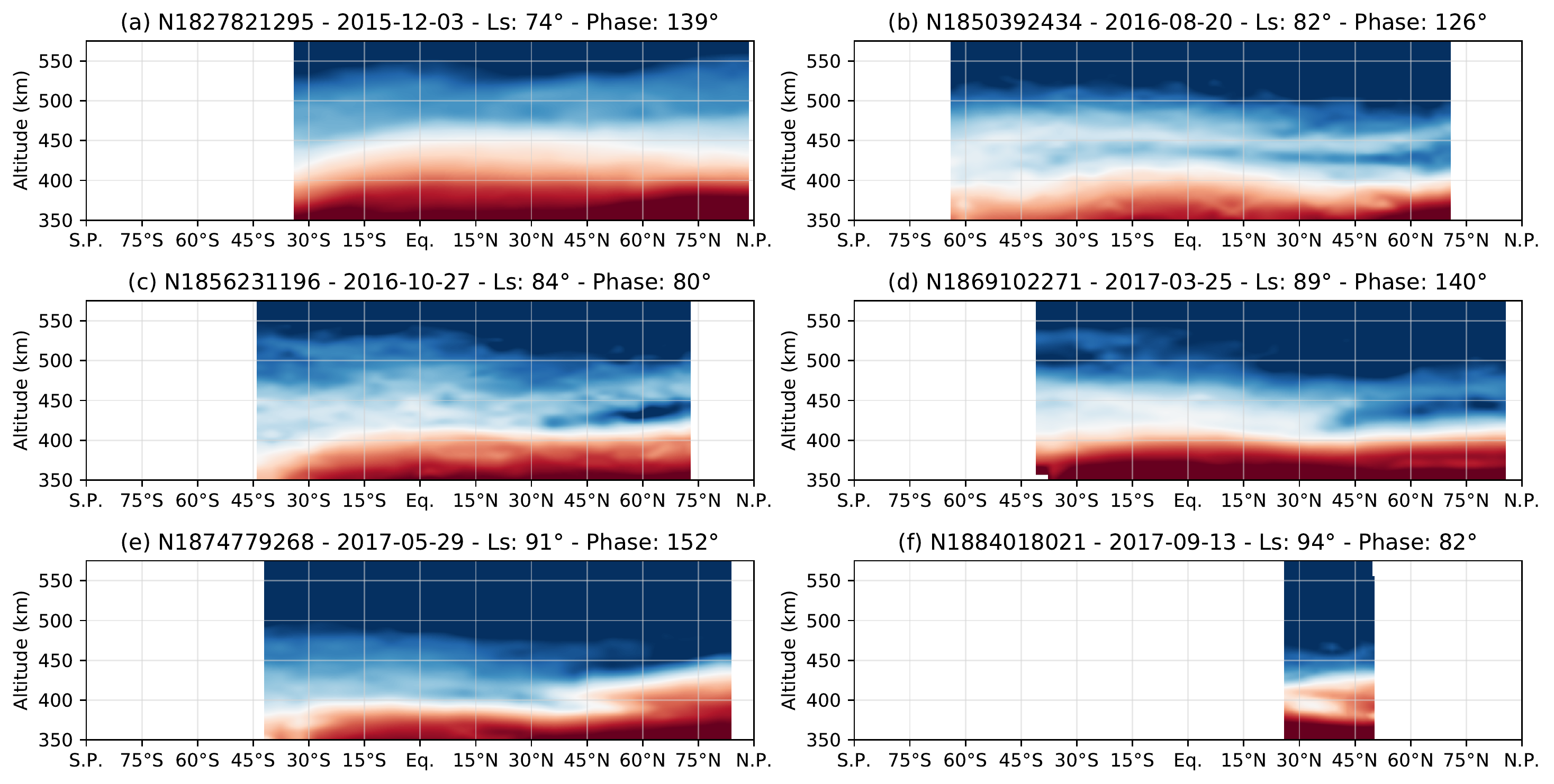}
\includegraphics[width=.5\textwidth]{Extinction_colorbar.pdf}
\caption{Same as the figures~\ref{fig:dhl_2004_2008}, \ref{fig:dhl_2008_2012}
and~\ref{fig:dhl_2012_2015} for 6 images taken between 2015 and 2017
($L_s=\ang{75}-\ang{95}$) during the reappearance of the DHL.
The image \textbf{N1884018021\_1} is one of the very last observations of Titan before
the end of the Cassini mission in September 2017.}
\label{fig:dhl_2015_2017}
\end{figure*}

The first occurrence of the stable detached haze in this period was seen on 3$^{rd}$ December 2015
(Fig.~\ref{fig:dhl_2015_2017}a). It
does not at first appear different from the previous sporadic detached haze layers observed in the period 2012-2015. Its signature is very weak, but, after
this observation, the detached haze was more pronounced and present in each following observation. The detached haze became stable in time,
similar to the detached haze before the equinox. Therefore, we consider this date as a limit to the beginning of the
reappearance of the detached haze ($L_s=\ang{74}$). The evolution of the haze during this period is displayed in
Fig.~\ref{fig:dhl_2015_2017}. These observations validate the long awaited reappearance of the detached haze layer,
just before the end of the Cassini mission in September 2017.

The detached haze layer is not very well defined but it can be perceived at all latitudes
around 490 - 520 km (Fig.~\ref{fig:dhl_2015_2017}a). We notice a contraction
the main haze: the top of the haze dropped by 50 km in January 2016 compared to October 2015
(Fig.~\ref{fig:dhl_2012_2015}d). The main elements of the reappearance (altitude and date) confirm the
predictions made by the general circulation models \citep{Lebonnois2012,Larson2015} and the prediction
reported in \cite{West2011}. The detailed comparison between the observations and the GCM is discussed further
in section~\ref{seq:comparisons}.

With time, the detached haze layer became more distinct and the zone of depletion more pronounced, especially in the
northern hemisphere (Fig.~\ref{fig:dhl_2015_2017}b). The situation seems to be analogous to an early stage of
the structure observed in 2004 (Fig~\ref{fig:dhl_2004_2008}a, but in the opposite hemisphere).
However, although the detached haze persists in time,
it also settles and almost merges with the main haze in October, 2016 (Fig.~\ref{fig:dhl_2015_2017}c).
In the  following observations (Figs.~\ref{fig:dhl_2015_2017}d and e), we are
witnessing the complex evolution of the newly-formed detached haze that merged with the main layer southward
of \ang{35}N while it remained stable around 450 km northward of \ang{35}N and seems to vanish in May, 2017 rather than settle.
This structure was still observed in the very last image of Titan taken by Cassini just before its final plunge into Saturn
 (Fig.~\ref{fig:dhl_2015_2017}f) in September, 2017.

A secondary detached haze layer appeared in October, 2016 in the southern hemisphere at high altitude (around 520 km in Fig.~\ref{fig:dhl_2015_2017}c). Its northern boundary is not well defined. This new structure will be persistent with
time, at planetary scale, up to the end of the Cassini mission but it gradually descended. The results reported by \cite{West2018}
concern the detached haze at the equator only. Although they already revealed a complex behavior of the detached haze layer, the
present observations show a dichotomy between the two hemispheres. The detail of the evolution, the split in a double layer
structure, and the formation and disappearance of several structures were completely unexpected. According to the GCMs mentioned earlier, six years
after equinox, the post-equinoctial circulation was supposed to be already installed with a planetary-scale circulation cell
from the southern hemisphere to the north polar region. Apparently, this is not the case in the observations.

Cassini observations from 2004 to 2017 do not completely cover half a Titan year. The first and last observations
were taken almost at the opposite season, $L_s=\ang{297}$ and \ang{94} respectively (\emph{i.e.} \ang{157} apart).
This prevents direct comparisons of the detached haze at opposite seasonal phases,
although both 2004 and 2017 images are taken more than a season after the previous solstice.

\section{Local and short-term variability of the detached haze layer}
\label{seq:local_variations}

So far, we have considered the evolution of the detached haze layer in the frame of the seasonal change.
We then discussed the long term-evolution at the planetary scale as a function of latitude.
In this section we consider sets of observations to characterize short-term and local behavior of the detached haze.
These characterizations could only be conducted on a limited number of observations and require very specific acquisition geometries. They allow us to observe localized, secondary order variations of the detached haze layer.
First, we choose several images taken a few hours apart to
evaluate the hourly variability of the detached haze. Next, we consider
observations at low phase angle which show simultaneously the two limbs of Titan. And, finally,
we consider observations taken from a near-polar point of view which can show
longitudinal variations in a narrow range of latitudes.

\subsection{Short time or spatial  variability}
\label{subseq:time_variations}

As presented before, we observed a large-amplitude variability of the
detached haze layer extinction profile at all latitudes below \ang{35}N (Fig.~\ref{fig:dhl_2004_2008}a
to \ref{fig:dhl_2004_2008}c) between December 2004 and June 2005. Fortunately, Cassini took in June, 2005 a series of 9 images of
Titan with a time-step of 80 minutes (Table.~\ref{tab:time_variability}).

\begin{deluxetable*}{ccccc}
\tablecaption{Sequence of 9 images of Titan taken June 4$^{th}$, 2005.
The longitude and local time are given for the profile at the equator
on the illuminated side of Titan.}
\label{tab:time_variability}
\tablewidth{0pt}
\tablehead{
\colhead{Image ID} & \colhead{Time (UTC)} & \colhead{Phase} &
\colhead{Longitude (Eq)} & \colhead{Local Time (Eq)}
}
\startdata
N1496548825\_1 & 03:32 & \ang{10.4} & \ang{10.7}W & 17:12 \\
N1496552665\_1 & 04:36 & \ang{10.3} & \ang{10.4}W & 17:17 \\
N1496557465\_1 & 05:56 & \ang{10.1} & \ang{10.6}W & 17:23 \\
N1496562265\_1 & 07:16 &  \ang{9.9} & \ang{13.1}W & 17:17 \\
N1496567065\_1 & 08:36 &  \ang{9.8} & \ang{13.4}W & 17:23 \\
N1496571865\_1 & 09:56 &  \ang{9.8} & \ang{13.9}W & 17:23 \\
N1496576665\_1 & 11:16 &  \ang{9.8} & \ang{14.3}W & 17:30 \\
N1496581465\_1 & 12:36 &  \ang{9.9} & \ang{14.7}W & 17:30 \\
N1496586265\_1 & 13:56 & \ang{10.1} & \ang{15.2}W & 17:36 \\
\enddata
\end{deluxetable*}

With multiple observations of Titan in a short period, we are able to validate our calibration and
observe short time and local variability. Here, we analyze the sequence at three different
locations on the limb (\ang{40}S, \ang{0}, \ang{40}N). The 9 observations are made with phase
angles around \ang{10}, within an interval of \ang{0.6}. The limb longitude of the observations
varies  between \ang{10}W and \ang{15}W whereas the solar local time on Titan varies between
17:12 and 17:36.

\begin{figure*}[!ht]
\plotone{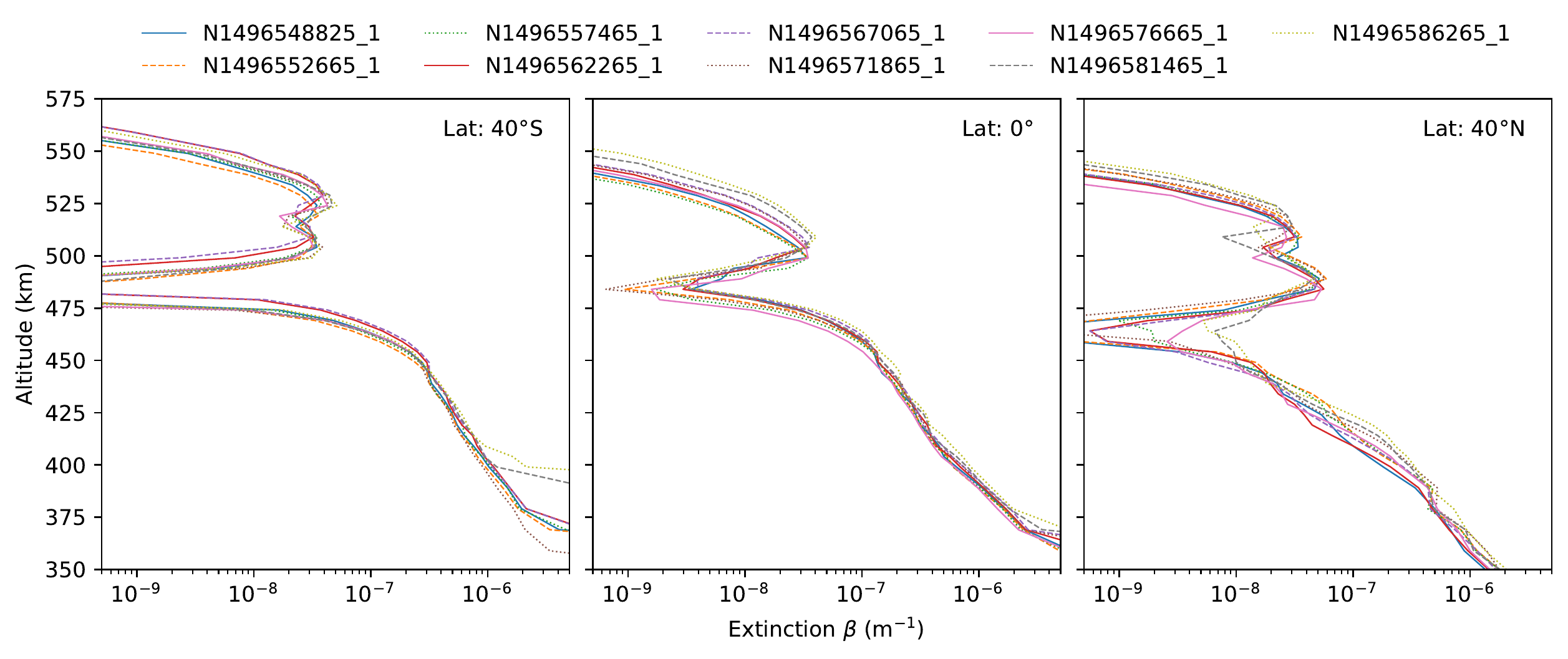}
\caption{Extinction profiles for the series
of 9 images taken 80 minutes apart in June 2005 (cf. Table \ref{tab:time_variability}).
The global latitudinal map is shown in figure~\ref{fig:dhl_2004_2008}c.}
\label{fig:time_variability}
\end{figure*}

Figure~\ref{fig:time_variability} presents extinction profiles extracted from the analysis of these 9 images.
First, we confirm that our calibration method is reliable from one image to another and the overall
variability of the detached haze is very small. We observe the double layering at \ang{40}S and \ang{40}N reported
previously. All the locations of the maximum of the extinction peaks are located within a small altitude range (smaller
than the 8 km of the pixel scale). Then, the vertical offsets which are observed at different times are consistent
with  the accuracy of the image navigation, and therefore are not significant. At the equator and at \ang{40}S, the detached
haze layer is well detached for all the profiles. When the vertical offset is accounted for, all the extinctions are found
with relative differences of about $\pm$ 10\% compared to the average value, except in the depletion zone between
the main and detached haze layers where difference can reach an order of magnitude (at \ang{0}) to several orders of magnitude
(\ang{40}N). The variations outside the depletion zone are not significant.

The variability in the depletion zone at the equator are about one order of magnitude and
consistent with uncertainties reported in the figure~\ref{fig:model_uncertainties}.
The variability observed at \ang{40}N, in the depletion zone around 460 km are significant for two reasons.
First, the differences are much larger than the expected uncertainty in this zone. Second, the sequence shows
a gradual and consistent increase of extinction with time. If these differences were due to uncertainties in
the retrieval procedure, it would have rather given a chaotic evolution of the extinction with time.

Furthermore, the corresponding extinction map (Fig.~\ref{fig:dhl_2004_2008}c) shows that in the south,
the detached haze layer is well separated from the main haze layer by a well defined depletion zone. In the north, the depletion
zone is less well defined and the detached haze and the main haze are connected vertically by a residual haze.
The variation of this residual haze is the one reported in Fig.~\ref{fig:time_variability} at \ang{40}N.
We can not strictly determine if we are witnessing time or a spatial variations since both the time and the longitude
of the observations change simultaneously during the image sequence. We note that a rotation of \ang{5} in longitude
corresponds to a maximum shift of 250 km in distance (one tenth of Titan's radius), possibly consistent with a spatial
variation of the haze extinction.

\subsection{Dawn and dusk sides}

\begin{figure*}[!ht]
\plotone{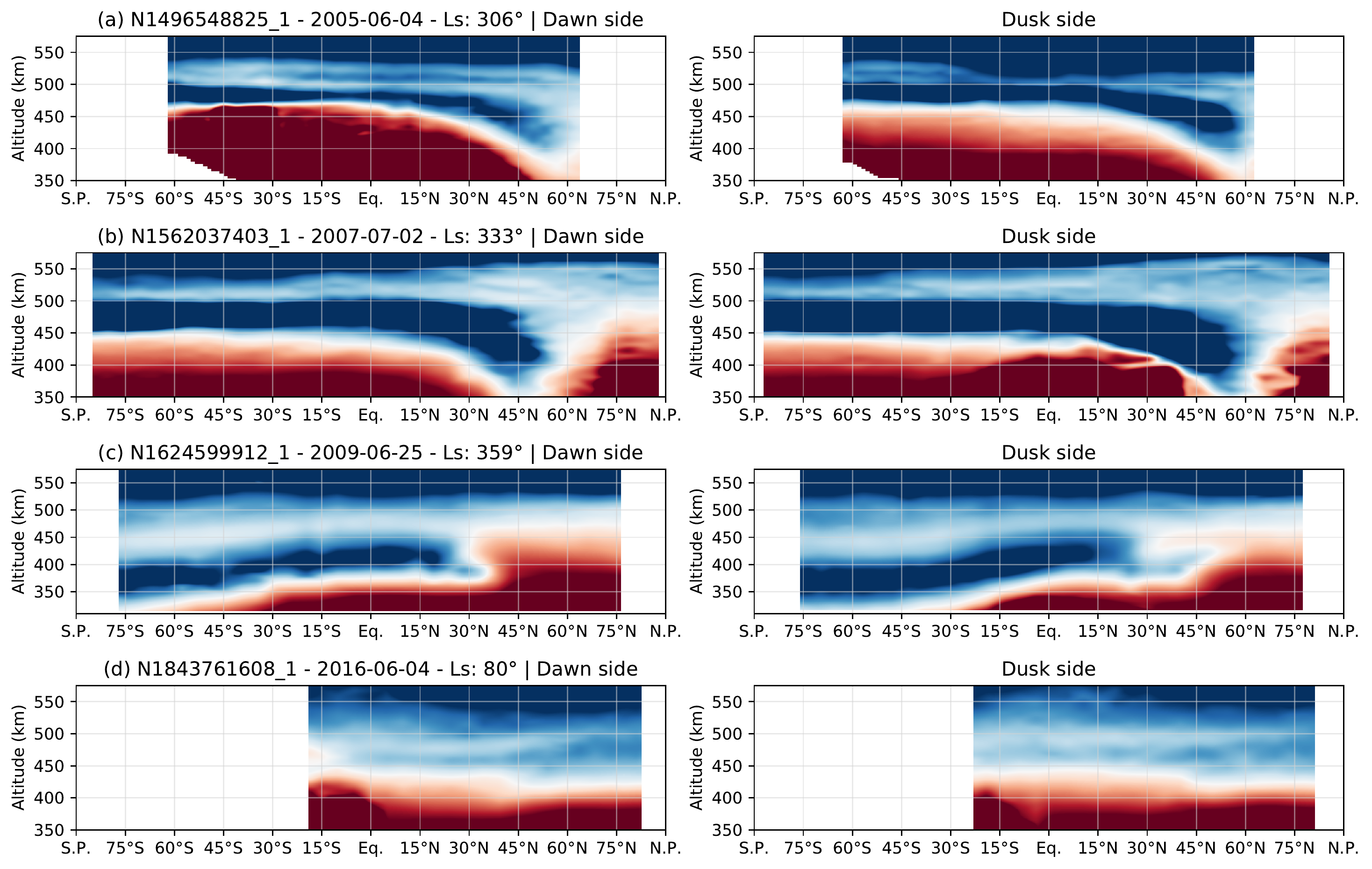}
\includegraphics[width=.5\textwidth]{Extinction_colorbar.pdf}
\caption{The left and right panels show maps of the haze extinction for the dawn and dusk side of
Titan for 4 different images taken during the mission when the viewing geometry allows us to observe both dawn and dusk
limbs. The date and the seasonal solar longitude are provided for each image.}
\label{fig:dawn_dusk}
\end{figure*}

Aside from short time and local variations, we also are interested in images showing simultaneously the two sides
of Titan. At low phase angle, the viewing geometry allows us to retrieve the haze extinction with both the
illuminated and the dark side of Titan simultaneously  (Fig.~\ref{fig:dawn_dusk}).
In this case, we can compare the dawn and dusk limbs for
specific latitudes. Although Titan's day is about 16 terrestrial
days, the time spent by the haze on the night side or dayside is much shorter.
First, the atmosphere is superrotating and at altitude around 400 or 500 km the zonal wind is comparable to or
larger than the rotation speed at the ground \citep{Flasar2005, Achterberg2011, Lebonnois2012, Lellouch2019}.
This makes the actual diurnal cycle for the high altitude hazes shorter than 16 days by a factor of 2 or more.
Secondly, at high altitude, sunlight penetrating beyond the geometric terminator further shortens the time spent in darkness.
Thus, effects on the haze should be produced by processes with timescales
comparable with a terrestrial day.

Figure~\ref{fig:dawn_dusk}a presents the haze extinction at the dawn and dusk sides as observed
in June of 2005 (Fig.~\ref{fig:dhl_2004_2008}c). The detached haze differs significantly between the two sides.
At the equator, we observe on the dawn side a double detached layer of 40 km thickness whereas it appears
as a thin layer of 10 km thickness on the dusk side. The haze extinction at the peak differs significantly,
from 7$\times 10^{-8}$ to 2.5$\times 10^{-8}$ m$^{-1}$, respectively. The depletion below the haze layer
is also less pronounced on the dawn side compared to the dusk side. Although this observation was taken
during the period of stability for the detached haze, we observed a significant asymmetry between
the dawn and dusk side. This asymmetry is also observed in all the images taken at the same moment which were
analyzed in section~\ref{subseq:time_variations}.

Two years later, another low phase angle image was recorded (Fig.~\ref{fig:dawn_dusk}b).
This time, the detached haze layer is much more symmetrical between the two sides. The peaks of haze extinction
are at the same altitude with comparable values. However, small differences can be noticed: there is a
small secondary layer above the detached haze layer at the dusk side between \ang{50}S and \ang{20}N, and may even extend
northward. In the northern hemisphere, the extinction appears slightly larger at dawn that at dusk, and the
vertical extent of the detached haze is also a bit larger. However, the overall morphologies are very similar.

At the equinox, the detached haze layer already started its drop in altitude (Fig.~\ref{fig:dawn_dusk}c).
There are significant differences between the two sides, in both hemispheres, while at the equator the two
profiles are almost identical. The haze layer is not exactly symmetrical in the southern hemisphere. The detached
haze itself is at the same altitude, but the depletion zone is at higher altitude at the dawn side, and the main
haze layer is thinner in the dusk side. In the northern hemisphere, the haze layer is more complex, and the
asymmetry is even more marked with a detached haze at different altitudes and with a different extinction. The
detached and main haze layers at the dawn side appear optically thicker than the layers at the dusk side.
This is consistent with the two previous cases.

After the equinox, fewer images were taken at low phase angle. Among them, we do not notice any
significant differences between the dawn and dusk sides. After the reappearance of the detached haze layer in
2016, we found only one image with the relevant geometry to see both sides of Titan illuminated at the same time
(Fig.~\ref{fig:dawn_dusk}d). In this case, we are close to the solstice and the sub-solar latitude is almost
at its maximal extent and does not allow us to probe the southern high latitudes near the terminator.
The main haze appears very symmetrical on both sides and almost identical
above \ang{45}N. The depletion can be followed continuously all around the North Pole. At latitudes lower than
\ang{45}N, the values of the haze extinction are similar but the altitudes of the extinction peak differ by
25 km.

The haze extinction profile and the altitude of the detached haze layer can differ between the dawn
and dusk sides. In general we notice a higher extinction in the dawn side than on the dusk side. This effect
could be due to a diurnal cycle.

\subsection{Longitudinal variability}

Due to orbital constrains, most of our observations sampled only a small range of longitudes.
However, a few observations of Titan taken from a near-polar point of view offer a unique way to study the evolution of
the detached haze with a large coverage in longitude and within a small range of latitudes. It allows us
first to check the homogeneity of the haze in longitude, and it also allows us to extend our previous
observations between the dawn/dusk sides with a local time coverage between 6:00 and 18:00.

\begin{figure*}[!ht]
\plotone{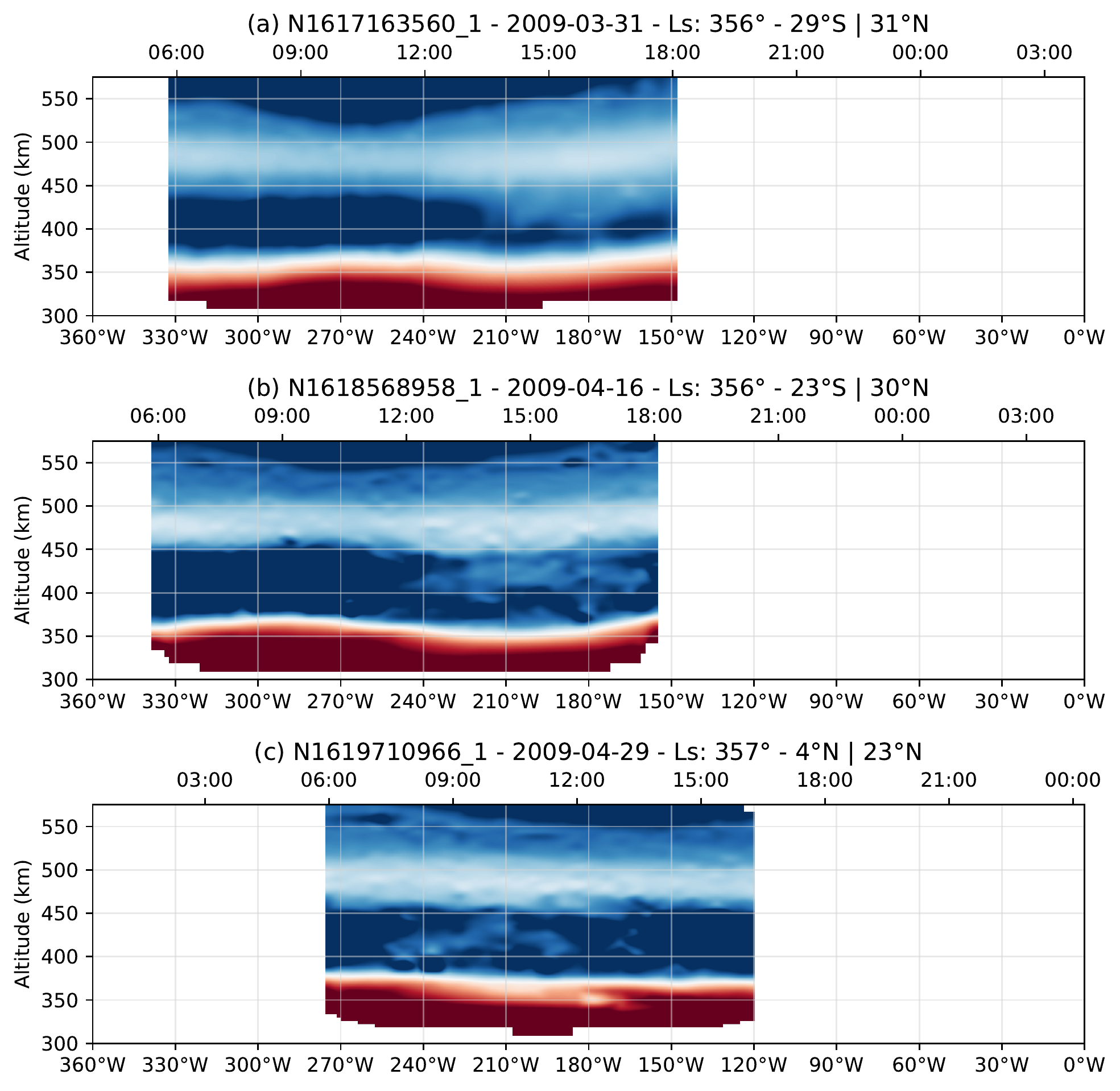}
\includegraphics[width=.5\textwidth]{Extinction_colorbar.pdf}
\caption{The panels show the map of the haze extinction as a function of longitude and local
time for a set of 3 images taken in March and April, 2009 ($L_s=\ang{356}$). The latitude range covered is
also indicated for each image.}
\label{fig:lon_variability}
\end{figure*}

We analyzed a set of three images taken sequentially within a month interval. The first observation
was performed on the 29$^{th}$ of March, 2009 (Fig.~\ref{fig:lon_variability}a) during the collapse of the detached
haze layer. Two other observations were made two weeks and one month later, with the same geometry
(Figs.~\ref{fig:lon_variability}b, \ref{fig:lon_variability}c). The detached haze layer is
located at 470 km at all the longitudes. Inside the depletion region, we notice a plume of haze
between 400 and 440 km and between \ang{150}W and \ang{220}W. In mid-April, the plume is located between
\ang{160}W and \ang{240}W and between 375 and 425 km. It appears disconnected and settling from the detached
layer, which remains at 470 km. At the end of April, we see the extension around 410 km and it
is spread from \ang{180}W and \ang{250}W. This aerosol plume is almost connected with the main haze.

This feature is not correlated with the local time but remains about at the same longitude and drifts slowly
toward the West. This would correspond to a retrograde motion of about 0.6 m/s. Another solution would be
prograde motion of 6.6 m/s, in phase with the sampling of 15 terrestrial days. The vertical speed, assuming
that the aerosol cloud dropped from 400 km to 375 km in one month would be $10^{-3}$ m/s. We also identified a
modulation in the extinction and in the geometrical thickness of both the detached haze layer and the main
haze. In the last image, only the geometrical thickness is modulated and not the haze extinction.

These observations show that the haze layer is not completely homogeneous in longitude and have some
fluctuations in extinction and in geometry.
It also strengthen the idea that space and time variations, as in observations previously discussed,
can not be distinguished without additional observations (not available in this dataset).
The dawn/dusk differences and the short-term variations,
presented in the two previous subsections, could be due to longitudinal effect rather than to time variations.
Therefore, with the results of this section, we stress that the longitude inhomogeneities should be kept
in mind as a secondary effect when discussing and comparing latitudinal maps of the detached haze layer.

\section{Comparisons with UVIS occultations and GCM predictions}
\label{seq:comparisons}

\cite{West2018} already confirmed the excellent agreement between the observations made during
two Voyager flybys and the position of the detached haze layer one Titan year later.
Comparisons with Cassini Visual and Infrared Mapping Spectrometer (VIMS) and
Cassini Composite Infrared Spectrometer (CIRS) instruments, could be possible but
they are limited by the sensitivity of their detectors above 450 km where the detached haze is located.
Therefore, we only performed a comparison with two stellar occultations
made by the Cassini Ultraviolet Imaging Spectrograph (UVIS) instrument in 2009 \citep{Koskinen2011}.
The extinction profiles retrieved in the previous sections can also be compared with results
obtained with other instruments and with Global Circulation Model (GCM) predictions.

\subsection{Comparison with UVIS occultations}

\cite{Koskinen2011} derived information on the mesosphere and thermosphere of Titan using UVIS stellar
occultations. The sensitivity to haze opacity of UVIS during a stellar occultation is much better than what ISS can achieve. However,
while ISS probes the light scattered by the detached haze layer, UVIS probes the light transmitted through a tangential
path at the limb. In both cases, an extinction profile can be retrieved. ISS can retrieve the extinction of the
particles which scatter light, and under assumptions concerning the phase function and the single scattering albedo \citep{Seignovert2017, West2018}.
On the other hand, UVIS is able to retrieve the total extinction from transmission with no assumption about the haze
particles. This difference is valuable because it may give information about the change in aerosol
size with altitude. In practice, ISS sensitivity is not sufficiently sensitive to probe above
the peak of the detached haze by more than a scale height.

\begin{figure*}[!ht]
\plotone{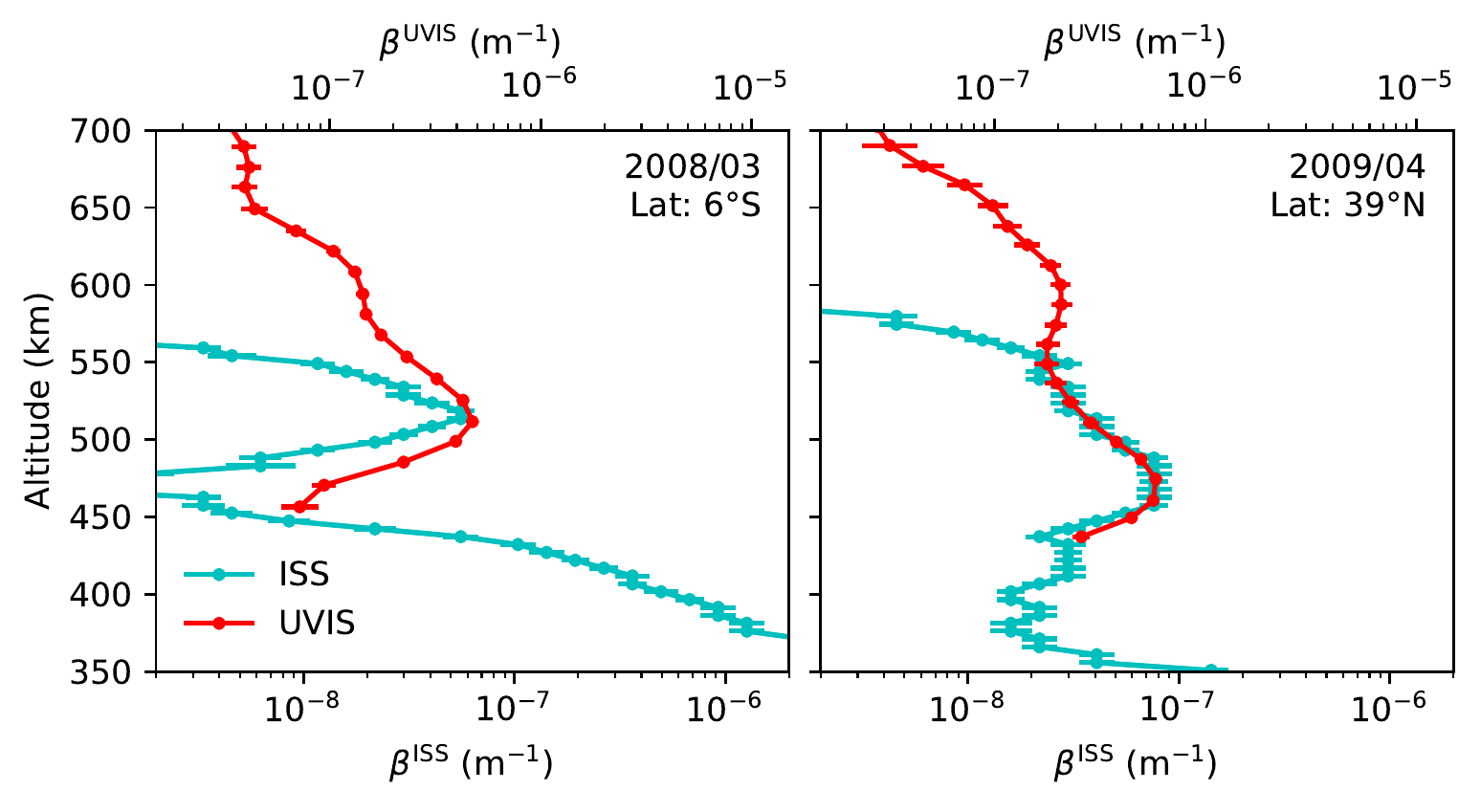}
\caption{Comparisons between ISS and UVIS extinction profiles before the equinox.
UVIS profiles are retrieved by \cite{Koskinen2011} during the T41 (2008/02/23) and T53 (2009/04/19) flybys.
ISS profiles are retrieved for the images N1585329510\_1 (2008/03/27) and
N1618568958\_1 (2009/04/16).
The UVIS profiles are scaled by a factor 0.15 to compensate the spectral dependence of the extinction
cross section and overlap ISS retrievals.}
\label{fig:uvis_iss}
\end{figure*}

Two of the UVIS occultation profiles, in 2008 and 2009 (T41 and T53 flybys), can be directly compared with ISS
observations at the same location and at the same period (Fig.~\ref{fig:uvis_iss}). The UVIS profiles are scaled to
offset the spectral dependence between ISS and UVIS effective wavelengths (338 nm and 1850-1900~\AA~respectively).
This offset is due to the spectral dependence of the extinction cross-sections and to the intrinsic
differences arising from comparing the extinction retrieved from scattering properties or from occultation
\citep[see.][]{Cours2011}.

The two profile can be compared in the 450 to 550 km altitude range.
In the first case (Fig.~\ref{fig:uvis_iss}a), even if the profiles don't exactly overlap, the ISS extinction profile presents a peak of extinction exactly at the same location as the one observed by UVIS. The drop, above and below the peak is more pronounced with ISS than with UVIS.
In the second case (Fig.~\ref{fig:uvis_iss}b), the two profiles presents a excellent agreement with each over in the 450 to 550 km altitude range.

Considering that UVIS and ISS profiles are not taken simultaneously and
they don't probe the same longitude, the results of the previous section demonstrate that these differences are
consistent with the natural variabilities observed in the detached haze layer.
This comparison is then a good validation of our results concerning the extinction profiles of the detached haze.

Above 575 km, UVIS extinction profiles show the presence of a secondary layer at 610 km which is not detected by ISS.
ISS is sensitive to the aerosols scattered throughout the limb, whereas UVIS observations probe the extinction along the tangential path.
Due to its stronger forward scattering peak, the large particles contribute more to the measured ISS scattering signal compared to the smaller particles, that scatter the light more isotropically. During an occultation, both small and large particles contribute to the UVIS extinction signal \citep{Cours2011}.
In theory, the difference between UVIS and ISS above 575 km could reveal a change in aerosol size distribution.
However, this layer is located at altitudes where the signal to noise is low where our model is no longer able to retrieve the extinction properly.
Therefore, it is not possible to draw a safe conclusion from the comparison between ISS and UVIS profiles above 575 km.

\subsection{Comparison with general circulation model predictions}

\begin{figure*}[!ht]
\plottwo{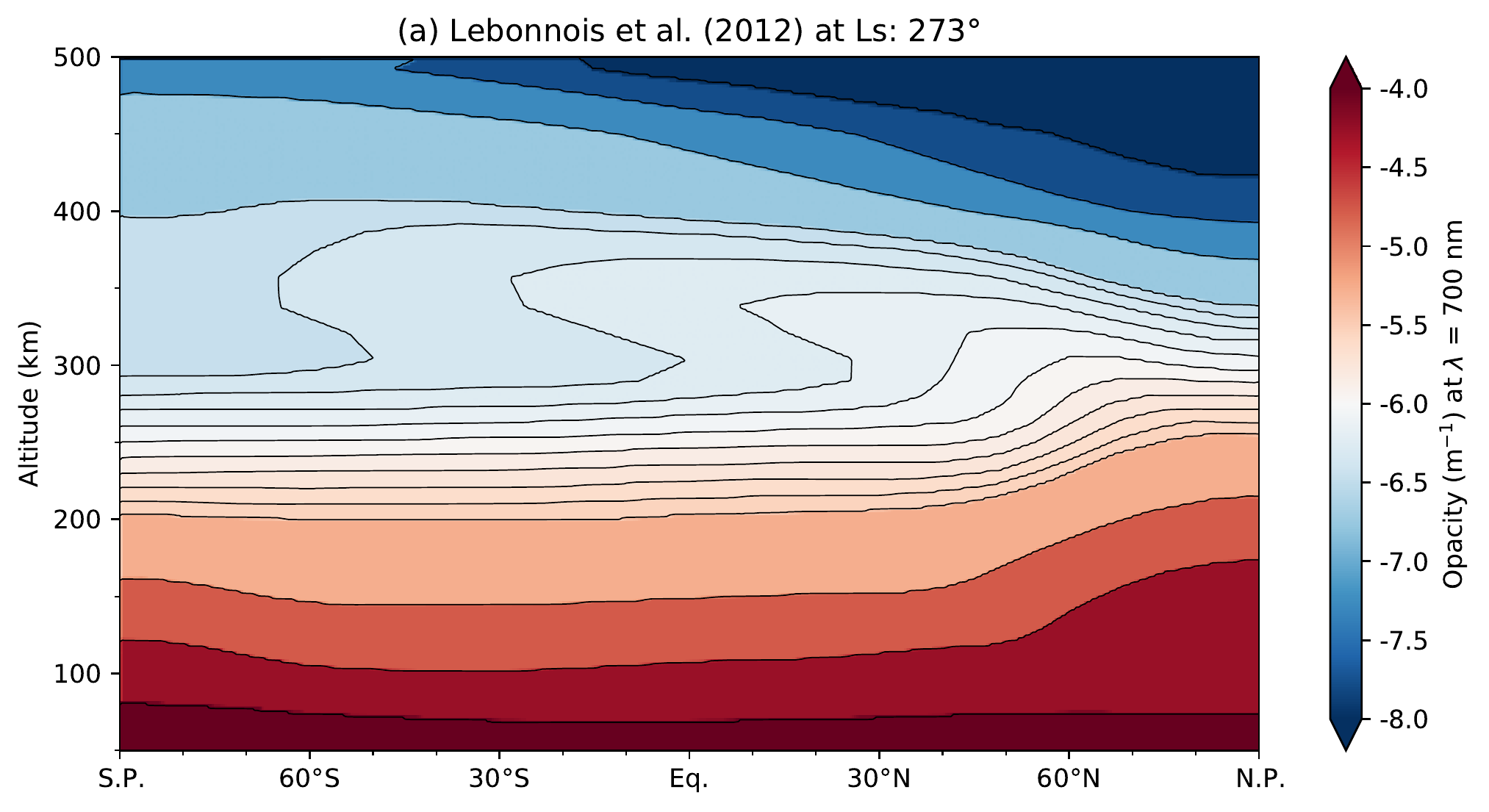}{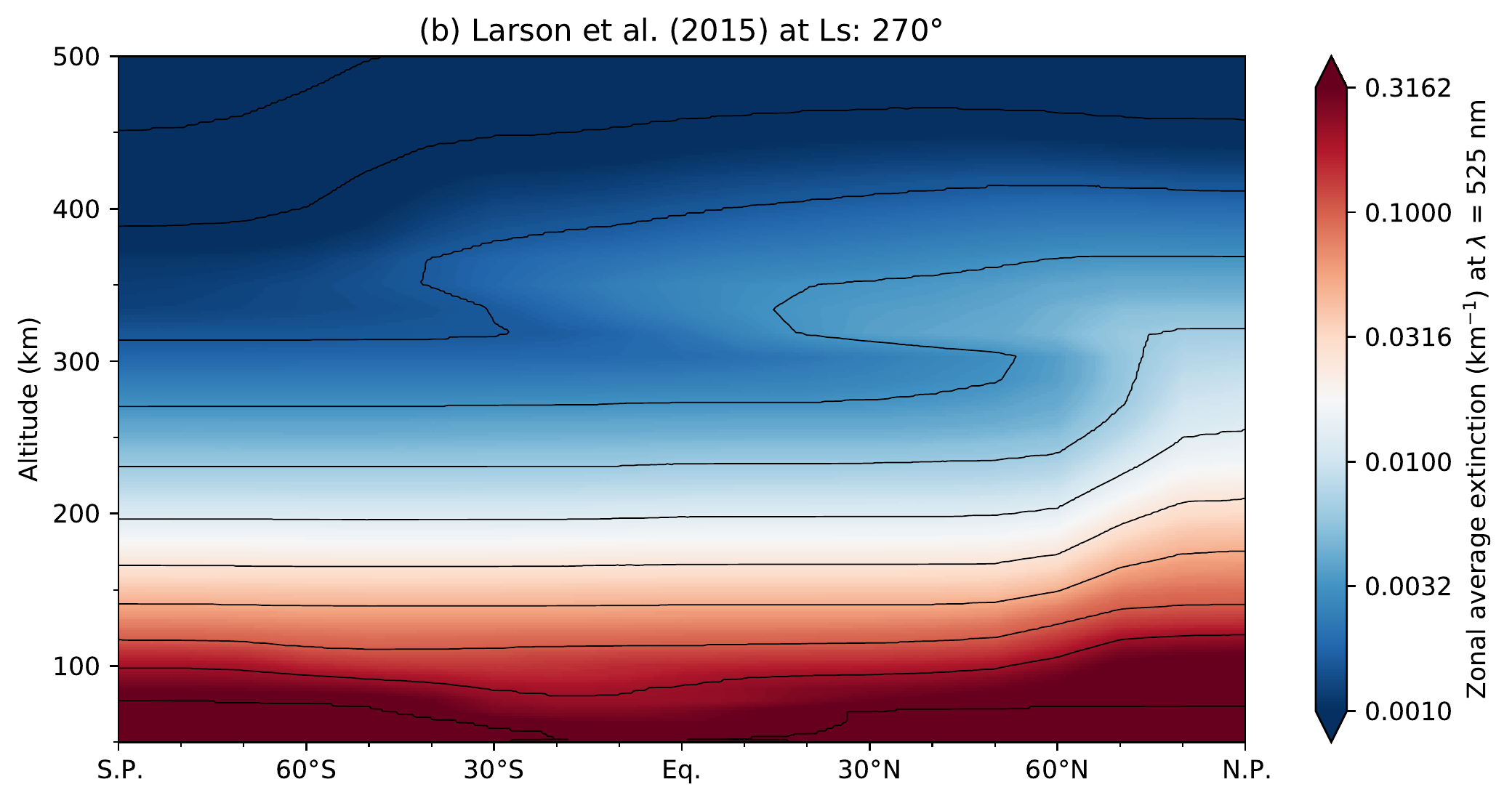}
\plotone{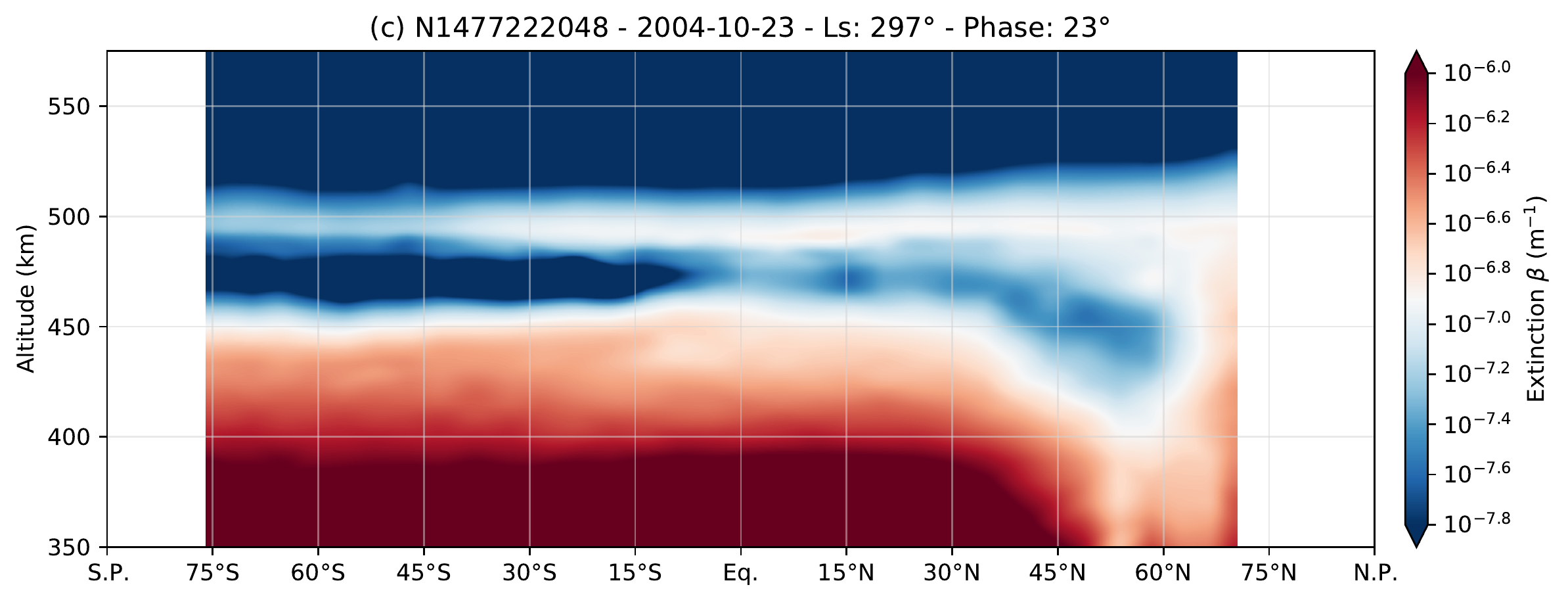}
\caption{Zonally-averaged haze extinction at northern winter solstice
($L_s = \ang{270}$) estimated by \cite{Lebonnois2012} at the wavelength $\lambda = $ 700 nm (a)
and by \cite{Larson2015} at $\lambda = $ 525 nm (b). (c) Haze extinction map
retrieved from Cassini/ISS observation CL1-UV3 ($\lambda = $ 338 nm) in the middle of winter
(N1477222048\_2 - $L_s = \ang{300}$).}
\label{fig:gcm_winter}
\end{figure*}

\begin{figure*}[!ht]
\plottwo{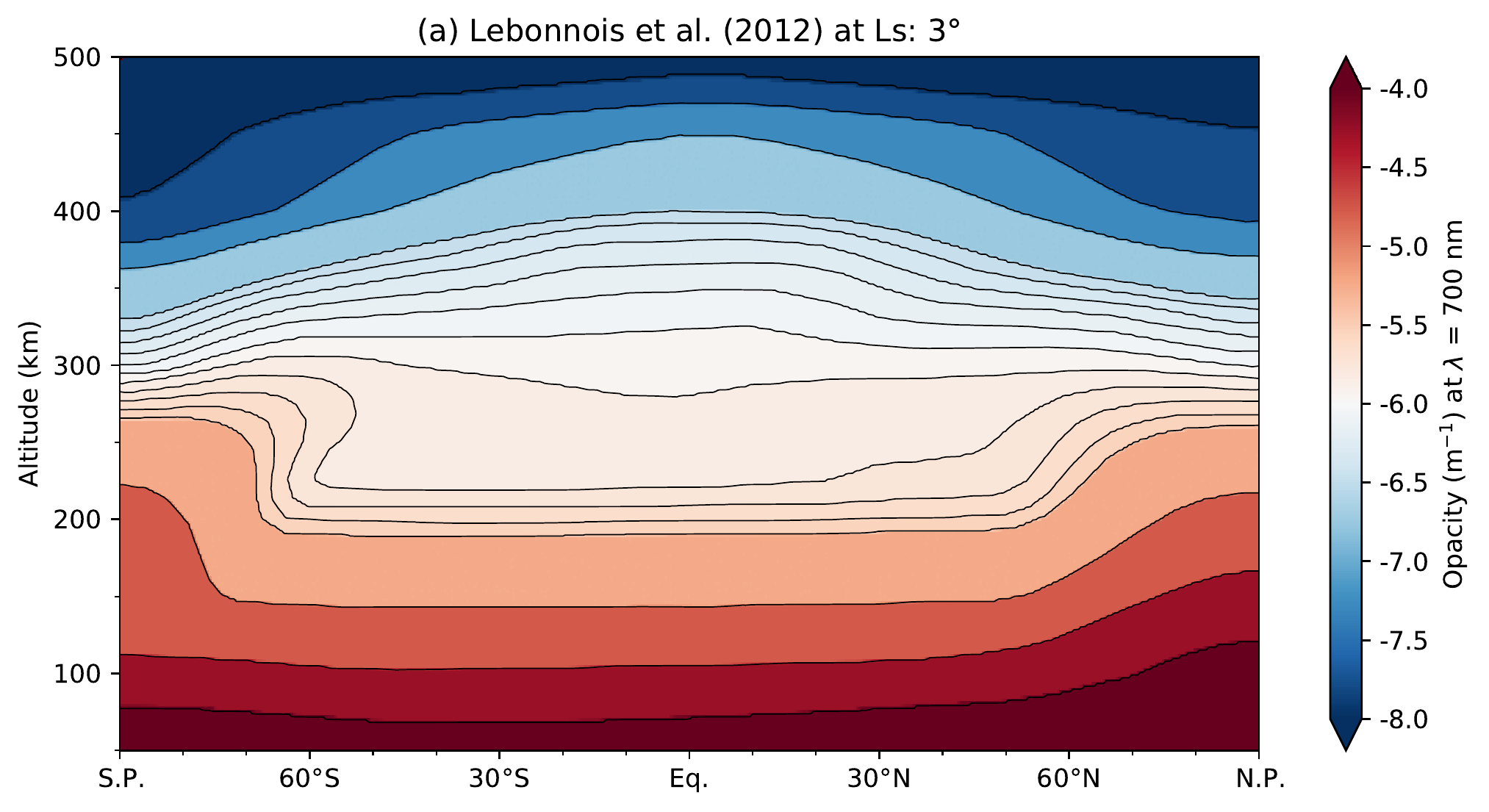}{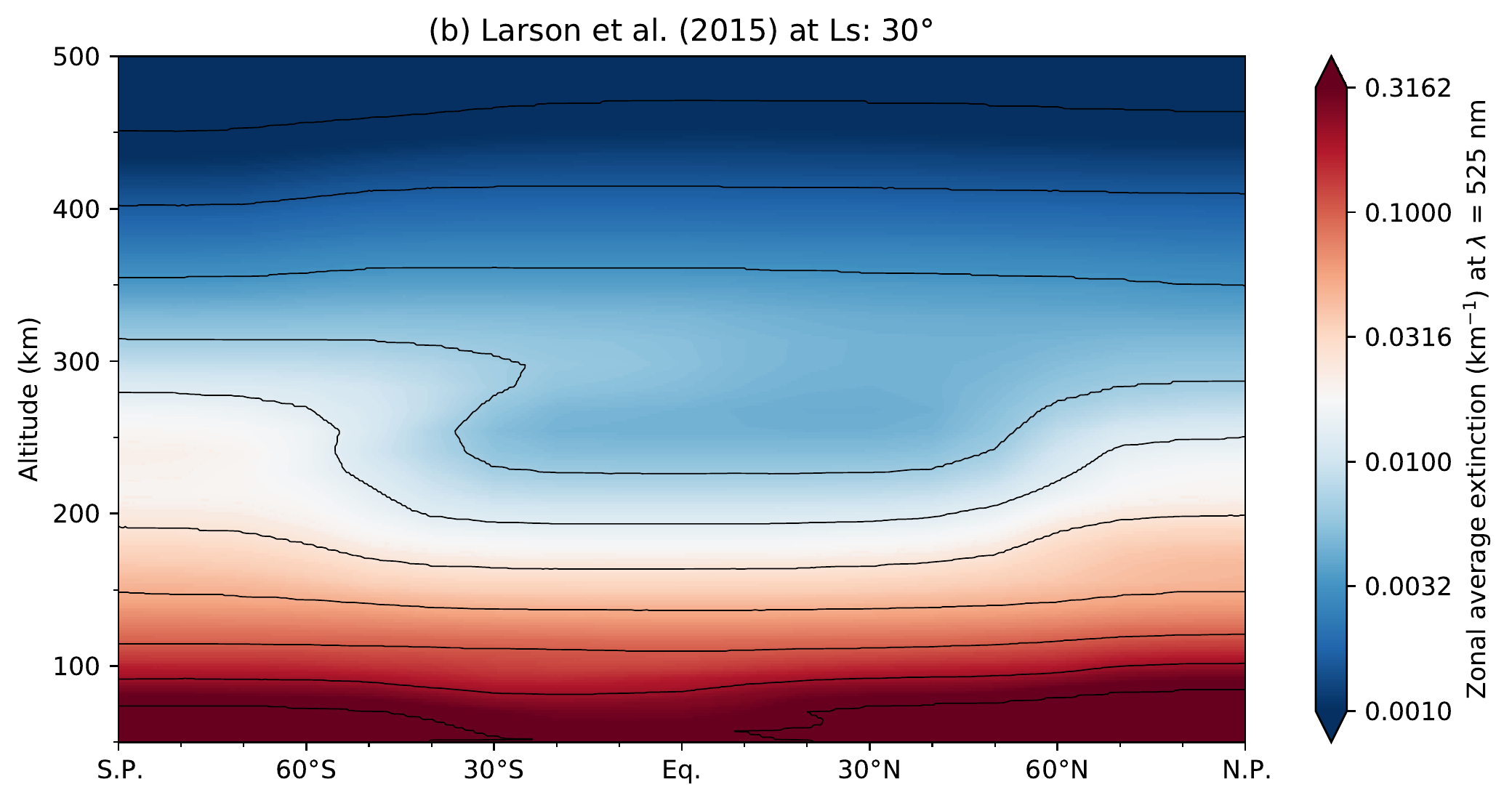}
\plottwo{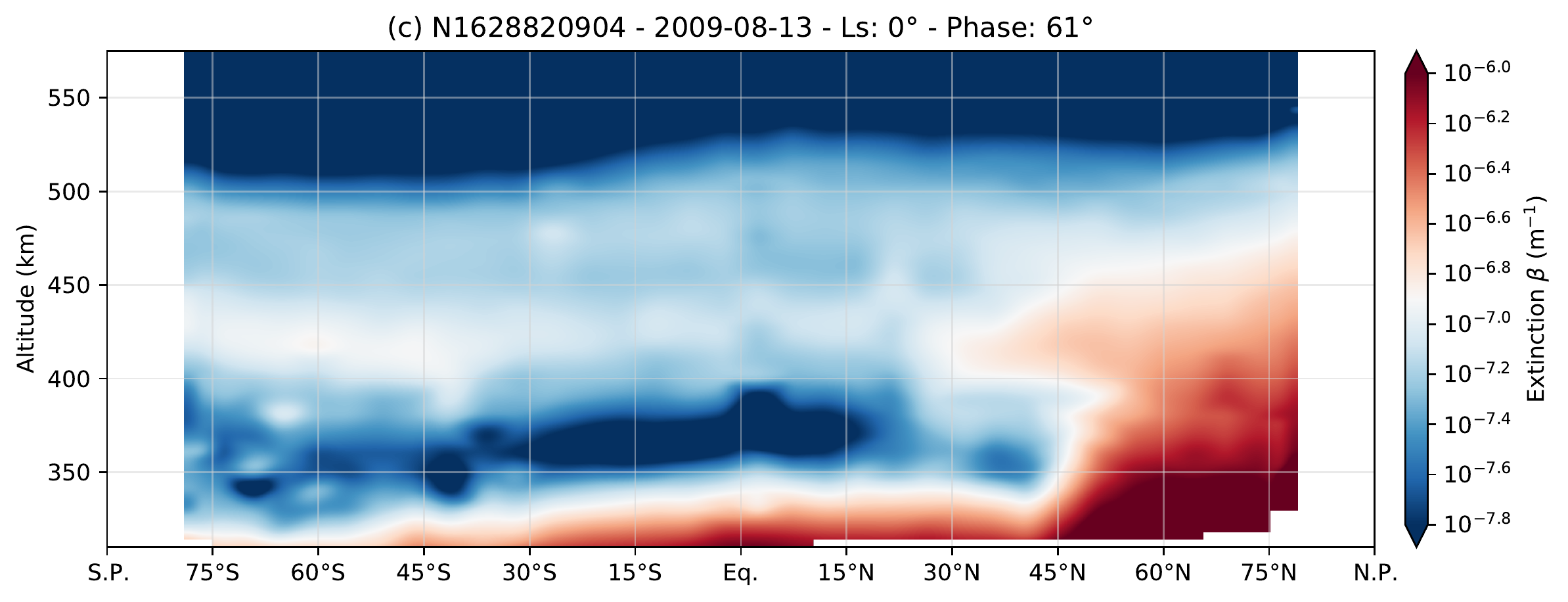}{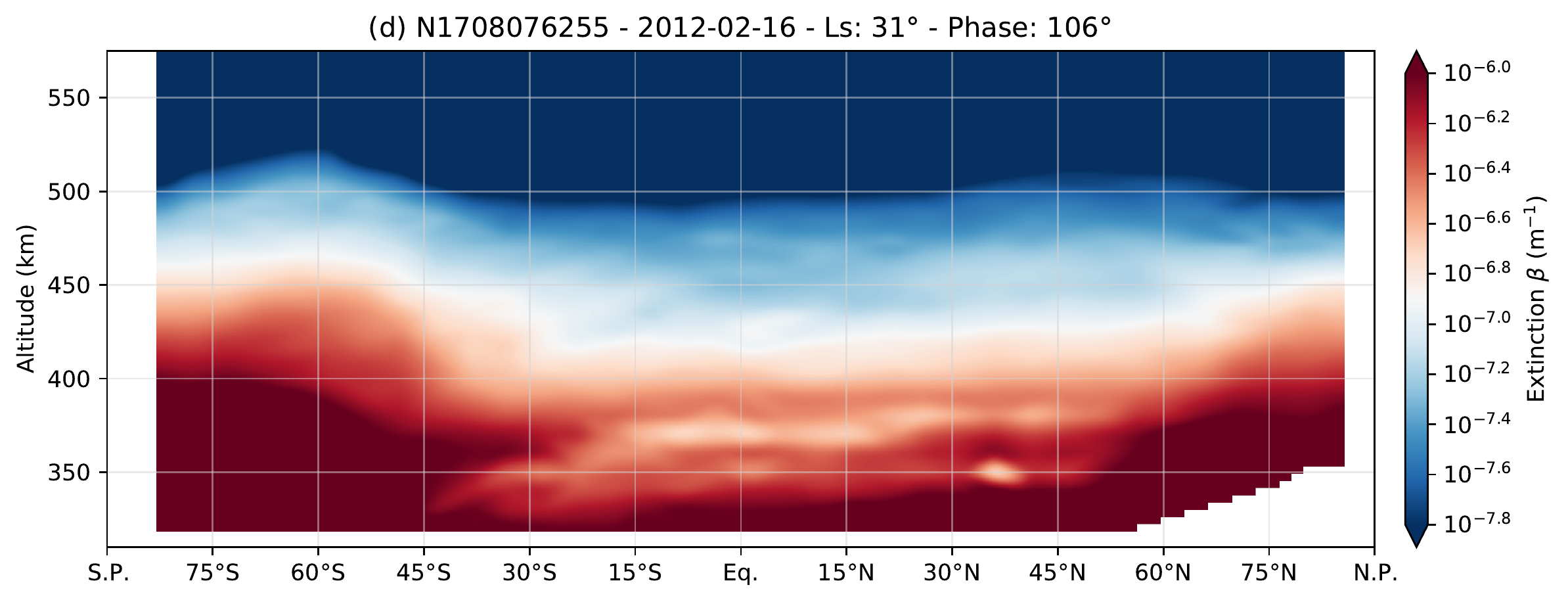}
\caption{Rescaled zonally averaged haze extinction at the northern spring equinox ($L_s = \ang{3}$)
estimated by \cite{Lebonnois2012} (a) and at 1000 days after the equinox ($L_s = \ang{30}$)
by \cite{Larson2015} (b).
(c-d) Haze extinction map retrieved from the Cassini/ISS observations
at the Northern Spring equinox (N1628820904\_1 - $L_s = \ang{0}$)
and 1000 days after the equinox (N1708076255\_1 - $L_s = \ang{30}$).}
\label{fig:gcm_spring}
\end{figure*}

General circulation models are very powerful tools to understand the climate of planetary atmospheres and the
interplay between different processes at planetary scales. In the case of Titan, circulation and haze are linked
by a strong feedback loop. The large-scale structures in the haze layer are produced by the action of the
circulation. The haze layer produces a feedback effect on the circulation through the control of the stratospheric
thermal structure \citep{Rannou2004}. The detached haze is one of the noticeable features produced by the the
stratospheric circulation \citep{Rannou2002, Lebonnois2012, Larson2015}. Figures.~\ref{fig:gcm_winter}
and \ref{fig:gcm_spring} show the maps of haze extinction obtained by \cite{Lebonnois2012} and
\cite{Larson2015} at 700 nm and 525 nm respectively.
Similarly to the comparison with UVIS, the distribution of the haze from the GCMs
can be compared with the extinction map derived with ISS in the CL1-UV3 filters at 338 nm with a scaling factor.

At the northern winter solstice (Fig.~\ref{fig:gcm_winter}), the detached haze appears around 350 km in both models.
In \cite{Lebonnois2012}, the altitude decreases by about few tens of km from the southern latitudes to
the north polar region where it merges with the north polar hood at \ang{40}N (Fig.~\ref{fig:gcm_winter}a).
In \cite{Larson2015}, the detached haze remains at constant altitude (Fig.~\ref{fig:gcm_winter}b), appears better marked than in \cite{Lebonnois2012}, and merges with the polar hood around \ang{60}N.
In both models, the extinction increases from the south to the north by about half a magnitude.
In the ISS observations made in 2004 (Fig.~\ref{fig:gcm_winter}c), \emph{i.e.} at the middle of winter, the detached haze layer is completely developed at 500 km and covers latitudes from the south polar region to \ang{60}N where it merges with the north polar hood.
The location of the depletion zone decreases from 475 to 425 km between \ang{40}N and \ang{60}N, which is not the case in
models. It is, on the other hand, consistent with the results obtained from stellar occultation by \cite{Sicardy2006}.
The haze extinction increases from the south to the north with about the same magnitude as in the models. This
is consistent with a layer increasing in aerosol loading while the airmass is flowing from south to north. It was already
noted \citep{West2011, West2018} that the detached haze layer in models appears as a supplementary layer added to
the background aerosols while, in data, it appears detached because there is a zone strongly depleted in
aerosols. Finally, as mentioned earlier, the detached haze layer is continuous all around the South Pole, which is not
the case in the models.

At the northern spring equinox (Fig.~\ref{fig:gcm_spring}), the model from \cite{Lebonnois2012} shows a
flat main haze layer without a detached haze between \ang{60}S and \ang{60}N and with two major increases at both poles (Fig.~\ref{fig:gcm_spring}a).
The haze in the south is increasing as a consequence of the circulation reversal while the northern haze is diminishing
and will disappear later in the season. In observations (Fig.~\ref{fig:gcm_spring}c),
this thicker haze is observed only for the northern latitudes, the detached haze has not yet disappeared
and the feeding of the south polar haze has not started. To notice a major increase of extinction at the South Pole,
we need to wait until the beginning of spring at $L_s = \ang{30}$ (Fig.~\ref{fig:gcm_spring}d).
At that time, the detached haze layer almost completely collapsed into the main haze and the haze distribution is very
similar to the one predicted by \cite{Lebonnois2012} at the equinox (Fig.~\ref{fig:gcm_spring}a).
In \cite{Larson2015}, 1000 days after the equinox (Fig.~\ref{fig:gcm_spring}b), we also observed a \emph{U} shape in the meridional haze extinction distribution as in data (Fig.~\ref{fig:gcm_spring}d), but in this case, a new detached haze layer already started to grow from the South Pole in the model (Fig.~\ref{fig:gcm_spring}b), whereas in data (Fig.~\ref{fig:gcm_spring}d) the local increase seen at 380 km is the consequence of the drop of a previous secondary layer (cf. Figs.~\ref{fig:dhl_2008_2012}g and \ref{fig:dhl_2008_2012}h).
In both comparisons, this means that the timing in the circulation models is not completely in phase with the observations.
The model of \cite{Lebonnois2012} seems to be in advance by about 3 years compared to the data.
We do not have enough information to characterize the advance in phase of \cite{Larson2015} model.

\begin{figure*}[!ht]
\plottwo{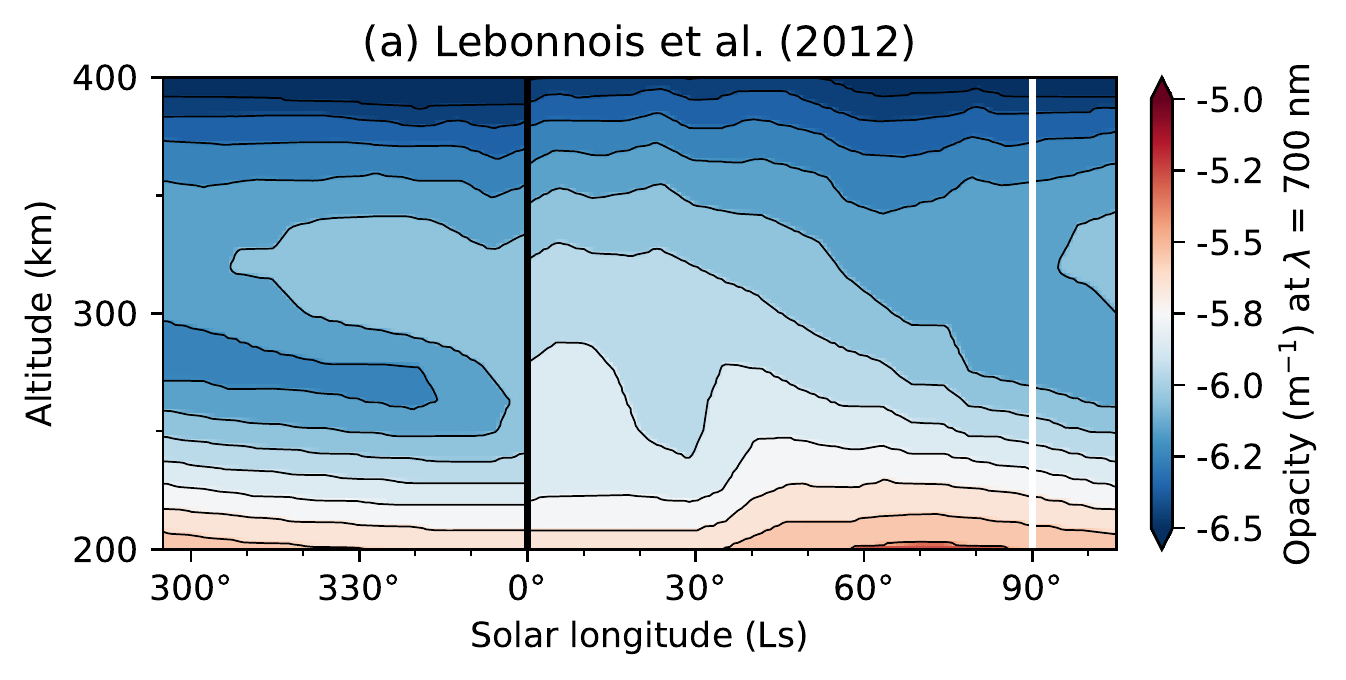}{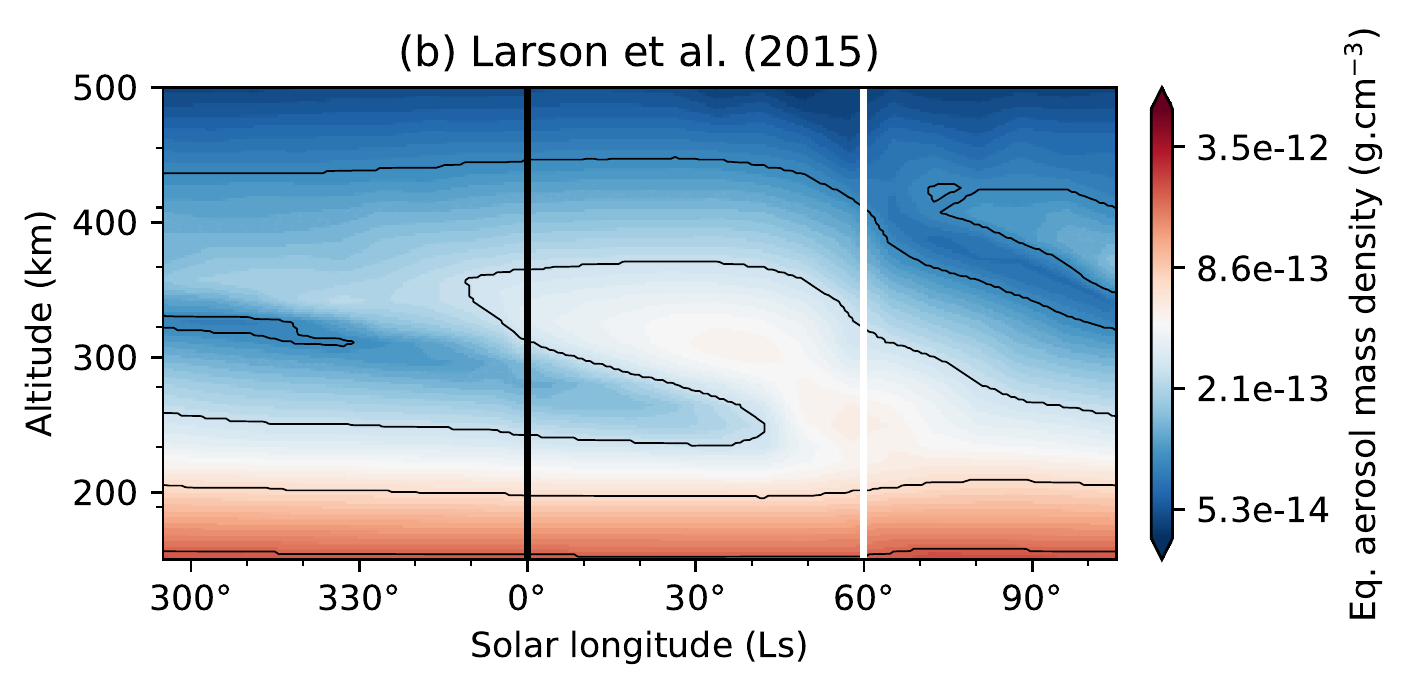}
\plotone{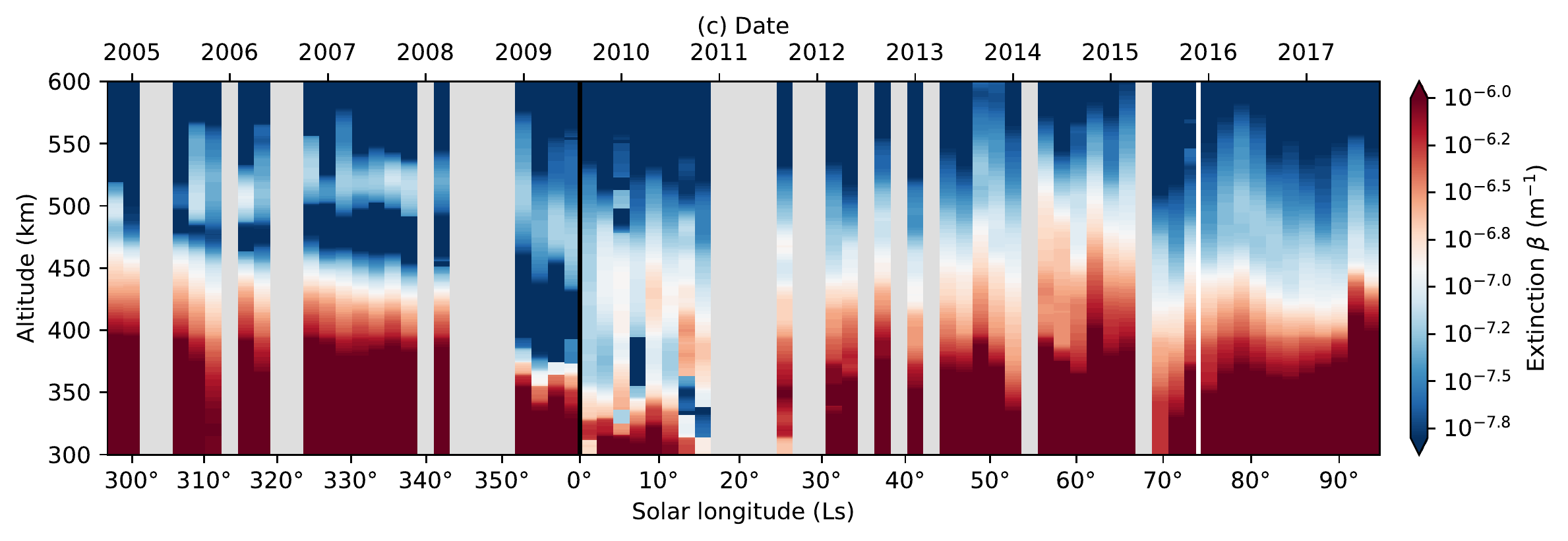}
\plotone{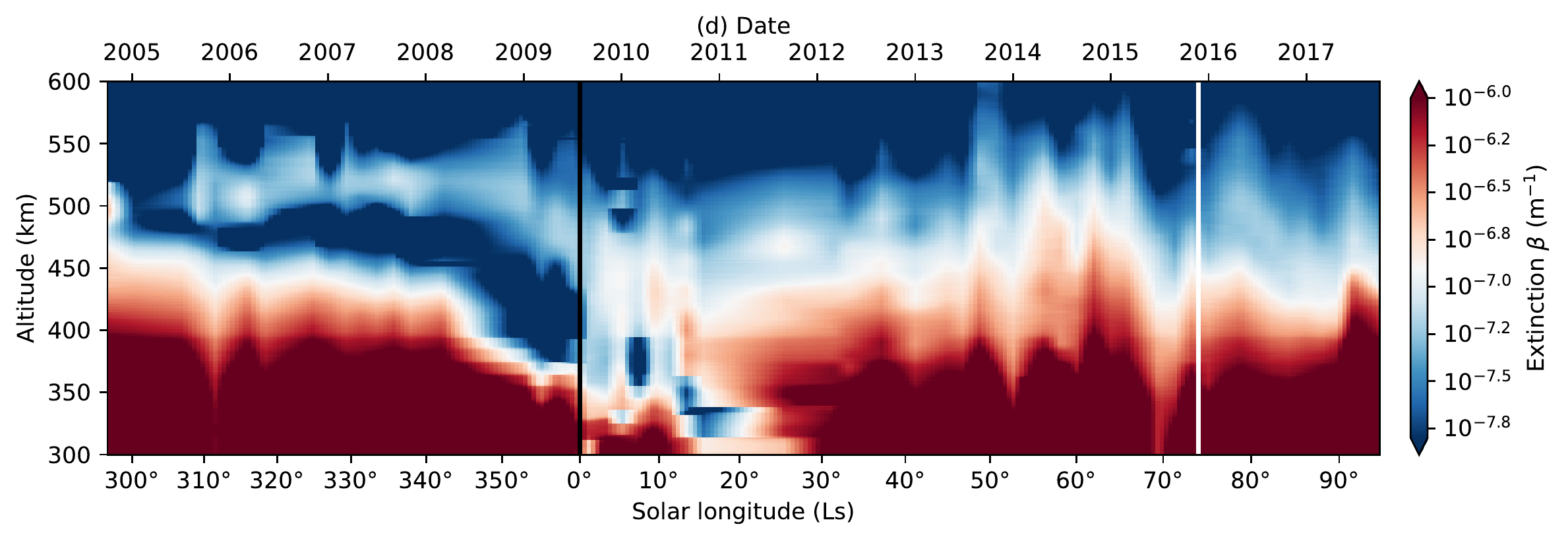}
\caption{(a) Annual variations of the zonally averaged equatorial opacity at 700 nm extracted from \cite{Lebonnois2012} and wrapped around $L_s = \ang{0}$.
(b) Seasonal evolution of the aerosol mass density (g/cm$^3$) adapted from \cite{Larson2015}.
(c) The haze extinction coefficient as a function of time and altitude at the equator (this work).
The data are binned by 2 months ($\Delta Ls \approx \ang{2} $). The gray area corresponds to the period where no UV3 observations where available (see Fig.~\ref{fig:img_sampling}).
(d) Same result as middle but linearly extrapolated in area without observations. This plot is only provided as an indicative global pattern but contains obvious interpolation artefact.
In each panel, the black vertical line corresponds to the vernal equinox ($L_s = \ang{0}$)
and the white vertical line is the date of the reappearance of the detached haze layer ($L_s = \ang{90}$, $L_s = \ang{60}$ and $L_s = \ang{74}$ respectively).}
\label{fig:gcm_cycle}
\end{figure*}

Anticipated in late 2014 or early 2015 ($L_s = \ang{60}$, Fig.~\ref{fig:gcm_cycle}a) by \cite{Larson2015} or in mid-2017 ($L_s = \ang{90}$, Fig.~\ref{fig:gcm_cycle}b) by \cite{Lebonnois2012}, the detached haze layer
finally reappeared in late 2015 to early 2016 ($L_s = \ang{74}$, Fig.~\ref{fig:gcm_cycle}c).
However, in May 2017, the upper atmosphere of Titan was still evolving and did not show a polar hood in the South Pole
similar to the one observed in 2004. Moreover, the most recent observations of early-2017 (Figs.~\ref{fig:dhl_2015_2017}c and \ref{fig:dhl_2015_2017}d) seem to show that the seasonal
formation of the detached haze layer could be different from one hemisphere to the other. The double peaks at
420 and 450 km in the temperature gradient profile, a proxy for the haze extinction, obtained from the 1989
occultation \citep{Sicardy1999} supports this hypothesis.

The current general circulation models do not match some of the large-scale features reported earlier. The first of them
is the shape of the vertical extinction profile which has a local depletion, producing the appearance of a detached haze layer in the
observations, while in models it appears as a high altitude haze layer superimposed to the main background haze.
The amplitude of the observed depletion reported, before and during the collapse, is sufficient to consider that the
detached haze layer is really disconnected from the main haze layer. We already stressed that the detached haze layer is
continuous around the South Pole without any visible upwelling coming from the main haze at this location. Finally, we
reported an early contraction of the main haze in 2008, just before the drop of the detached haze at the equinox. The origin
for such a contraction is likely related to the weakening of the Hadley cell when the latitudinal illumination gradient
decreases around the equinox.

Thanks to high spatial resolution of the ISS/NAC camera, we noticed some small-scale structures which could be unresolved or
erased by the temporal averaging in GCMs. During the northern winter and spring, we observed some sporadic decreases and
bursts in the extinction profiles at very short time scales. These events could have a major impact on the redistribution of
aerosols in the upper atmosphere. We also reported, in numerous cases, the existence of sub-layers above the main detached
haze layer with large latitude extend. Usually, their presence could be followed during more than one Earth year. This is
especially true during the collapse of the main detached haze layer after the equinox, when we observed several smaller
drops from 500 km down to 300 km. And finally, the double structure reported in early-2017
(Figs.~\ref{fig:dhl_2015_2017}c and \ref{fig:dhl_2015_2017}d) with two very distinct detached haze layers in the
North at 450 km, in the South at 520 km, was never mentioned before and not reported in the current GCMs. New modeling with a better vertical resolution will be need to interpret this structure.

We want to highlight facts that would help to better constrain the climate models.
GCMs are ideal tools to understand the interplay of different components of Titan's climate.
But they also have several limitations which are often technical by nature (low spatial resolution, shallow water
approximation for the dynamics, radiative transfer in plane-parallel, no link with the high mesosphere and thermosphere physics).
Limitations may also come from incorrect boundary conditions. We give some remarks:

\begin{enumerate}
\item the collapse starts in the southern hemisphere and concerns the main haze and then the detached haze. This
may indicate that aerosols of the main haze and the detached haze do not have the same fractal structure. This is
already suggested by \cite{Lavvas2009} and \cite{Larson2015}.
The weakening or the displacement of ascending branch of the circulation would first affect the more
compact aerosols of the main haze before the fluffiest aerosols of the detached haze that can remain suspended more
easily.
\item the finest details found in this work occur at length scales smaller than the GCM grid scale.
ISS observations show an instantaneous snapshot of the haze layer. GCMs generally
calculate extinction of haze averaged over several Titan days and/or are zonally averaged. Comparing GCMs instantaneous and
not zonally averaged outputs would allow to see if they are able or not to produce the same kind of variabilities as
observed. These variability may also be induced by processes occurring in higher layers not modeled by the GCMs.
\item the timing offset of the DHL reappearance in GCM is probably due to the model top levels which are generally too
low. Vertical extension of the models and of the haze layer may improve the timing of the cycle.
\item the source of the aerosols depends on photochemical processes that occur far above the DHL and which are
also are subjected to circulation in the upper mesosphere and thermosphere.
Accounting for the physics and chemistry in higher layers, for instance, a coupling with thermospheric models could
be a key to understand further the detail of the haze cycle.
\end{enumerate}

\section{Conclusions}
\label{seq:conclusions}

In this article we have analyzed observations of light scattered at the limb of Titan with ISS (UV3 filter) on the
Cassini spacecraft. We retrieved the haze extinction as a function of altitude, latitude and time between October 2004 and
September 2017. We followed, during about half Titan's year, the evolution of the DHL and the top of the
main haze. In particular, we witnessed the collapse of the DHL during the equinoctial transition of the atmospheric
circulation and its reappearance before the following solstice.

We confirmed and gave details about the collapse of the detached haze layer previously reported by \cite{West2011}.
We also tracked the small-scale variations after the equinox in order to detect the reappearance of the detached haze
layer at the end of 2015 \citep{West2018}. These two previous works focused on the DHL at equator. Here, we give
a full description of the structures of the haze layer as a function of altitude, latitude and time. We find
that the DHL has a natural variability, which can be temporal and spatial, and sometimes two distinct hazes or plumes
can be observed above the DHL. The amount of data is not large enough to distinguish between spatial or
longitudinal variability. But data taken with a polar viewing indicates that the haze is not completely
uniform in longitude.

The equinoctial collapse starts in the summer hemisphere. Its initial phase can be discerned in March, 2008.
The main haze collapses first and then the detached haze layer about one terrestrial year later. By April, 2012,
the detached haze is below 300 km and can not be seen at UV wavelengths.
The fall of the detached haze layer between 2009 and 2011 occurred at the average speed of -67 km/yr.
During the equinoctial collapse, the DHL seems to settle down at the aerosol terminal speed,
as reported by \cite{West2018}. If sedimentation does control the speed of the collapse,
it would indicate an absence of vertical wind at these two moments.
During a period of 3 years and a half (from mid-2011 to 2015), no stable detached haze layer was observed.
However, the haze layer fluctuates, and sporadic local detached layers appeared and disappeared rapidly.

The detached haze layer reappeared in December 2015. We first noticed this new detached haze when it was marginally
apparent in UV3 images.
The timing of the reappearance is offset compare to GCM predictions and it occurs with patterns more complex
that those predicted. First, it reappeared around 500 km in December, 2015 ($L_s =\ang{74}$) as a very faint
structure which became persistent and more pronounced with time.
At the equator this structure sedimented and finally disappeared in about one year, whereas it remained visible in the northern hemisphere.
A second detached haze appeared in July, 2016 around 500 km altitude, above the first DHL, and apparently started to settle down
as well. Unfortunately, the survey was interrupted in September 2017 by the end of Cassini mission. This second detached
haze layer did not cover all latitudes, was quite variable, and was present up to the final observation.

Unfortunately, no data were acquired between April 2008 and February 2009 and between December 2010 and September 2011.
We thought to use the NAC UV1 and UV2 images but these data have a very poor signal to noise ratio and
can not be included in our analysis.
However, a few images are available with the WAC camera in the VIO (Violet), BL1 (Blue) filters.
The behavior of the aerosols at these wavelength should be very similar to the UV3 filter and could fill these gaps.
It would also be interesting to perform similar analysis with data acquired through filters at larger wavelengths.
In these case, it would be possible to probe deeper layers in order to monitor the collapse of the DHL further
down and, as well, the cycle of the main haze. For instance, \cite{Rages1983} could probe as lower as 200-250 km
in clear filter ($\lambda_{eff} \simeq 0.5 \mu$m). At even longer wavelengths, we could reach levels in the low
stratosphere and, maybe probe high altitude polar clouds \citep{deKok2014,West2016}.

Comparison with General Circulation Models are fruitful. Our results reinforce, the scenario of a
detached haze cycle primarily controlled by circulation as proposed by \cite{Toon1992} with a 1D model, \cite{Rannou2002}
with a coupled 2D-GCM and \cite{Lebonnois2012, Larson2015} with coupled 3D-GCMs. Although GCMs capture the global
haze cycle, many differences remain, mainly driven by technical limitations.

\section*{Acknowledgments}

This work was supported by the French ministry of public research. The authors also thank the Programme National de Plan\'{e}tologie (PNP) for their financial support.
B.S. and R.W. thank the Cassini Mission, in particular the advocacy by Trina Ray and Mou Roy for obtaining the images studied here.
Part of this work was performed by the Jet Propulsion Laboratory, California Institute of Technology, under a contract with the National Aeronautics and Space Administration (80NM0018D0004).
P.R. and S.V. thank the French Agence Nationale de la Recherche (ANR Project APOSTIC No. 11BS56002, France).
B.S. thanks the Planetary Rings Node (SETI Institute) for their excellent PDS \href{https://tools.pds-rings.seti.org/opus}{browser interface OPUS}.

\section*{Data availability}
The list of all the ISS images and UV extinction profiles derived from this analysis, plus the source code to produce the figures, will be made
publicly available after publication on the \href{https://data.caltech.edu}{Caltech Data Archive},
doi:\href{https://doi.org/10.22002/d1.1344}{10.22002/d1.1344}.

\bibliography{biblio}
\bibliographystyle{aasjournal}

\end{document}